\documentclass[twocolumn]{aastex701}
\usepackage{amsmath}
\usepackage{verbatim}

\usepackage[T1]{fontenc}
\usepackage{hyperref}

\begin{document}

\title{Impact of Equation of State on Black Hole Accretion Flows and Radiative Properties}

\author[0009-0005-8944-2159]{Yun-Ming Zhu}
\affiliation{Tsung-Dao Lee Institute, Shanghai Jiao-Tong University, 1 Lisuo Road, Shanghai, 201210, People's Republic of China}
\email[show]{zhuyunming@sjtu.edu.cn}

\author[0000-0002-4064-0446]{Indu K. Dihingia}
\affiliation{Institute of Fundamental Physics and Quantum Technology, School of Physical Science and Technology, Ningbo University, Ningbo, Zhejiang 315211, P.R. China}
\affiliation{Zhejiang Key Laboratory of Extreme Universe, Ningbo, Zhejiang 315211, P.R. China}
\affiliation{Tsung-Dao Lee Institute, Shanghai Jiao-Tong University, 1 Lisuo Road, Shanghai, 201210, People's Republic of China}
\email{ikd4638@gmail.com, ikd4638@nbu.edu.cn}

\author[0000-0002-8131-6730]{Yosuke Mizuno}
\affiliation{Tsung-Dao Lee Institute, Shanghai Jiao-Tong University, 1 Lisuo Road, Shanghai, 201210, People's Republic of China}
\affiliation{School of Physics \& Astronomy, Shanghai Jiao-Tong University, 800 Dongchuan Road, Shanghai, 200240, People's Republic of China}
\affiliation{Key Laboratory for Particle Physics, Astrophysics and Cosmology, Shanghai Key Laboratory for Particle Physics and Cosmology, Shanghai Jiao-Tong University, 800 Dongchuan Road, Shanghai, 200240, People's Republic of China}
\affiliation{Institut f\"{u}r Theoretische Physik, Goethe Universit\"{a}t, Max-von-Laue-Str. 1, D-60438 Frankfurt am Main, Germany}
\email[show]{mizuno@sjtu.edu.cn}

\author{Ziri Younsi}
\affiliation{Mullard Space Science Laboratory, University College London, Holmbury St. Mary, Dorking, Surrey, RH5 6NT, UK}
\email{z.younsi@ucl.ac.uk}

\author{Christian M. Fromm}
\email{christian.fromm@uni-wuerzburg.de}
\affiliation{Institut f\"ur Theoretische Physik und Astrophysik, Universit\"at W\"urzburg, Emil-Fischer-Str. 31, D-97074 W\"urzburg, Germany}
\affiliation{Institut f\"{u}r Theoretische Physik, Goethe Universit\"{a}t, Max-von-Laue-Str. 1, D-60438 Frankfurt am Main, Germany}

\begin{abstract}
Previous literature has documented that black hole shadow images are mainly shaped by spacetime geometry and plasma microphysics governing accretion flows. However, the influence of equations of state (EoS) remains under-explored. 
In this paper, we quantify how different plasma microphysics prescriptions, specifically the choice of EoS and electron-heating models, affect flow thermodynamics and the corresponding synthetic black-hole images. We perform three-dimensional general-relativistic magnetohydrodynamics (GRMHD) simulations using two constant-$\gamma$ ideal EoSs with $\gamma = 4/3$ and $5/3$ and a temperature-dependent variable EoS (TM) in both accretion flow states: Standard And Normal Evolution (SANE) and Magnetically Arrested Disk (MAD) regimes. The dynamical models are post-processed with general-relativistic radiative transfer (GRRT) calculations at 86~GHz and 230~GHz, employing thermal and hybrid $\kappa$ electron distribution functions for synchrotron radiation. We also compare the two-temperature electron heating prescription based on turbulent heating and magnetic reconnection.

The constant-$\gamma$ EoSs systematically overestimate or underestimate gas and electron temperatures across disk and jet regions, whereas the variable EoS  provides a smoother trans-relativistic interpolation between the two limiting regimes. These differences substantively impact the synchrotron emissivity, image morphology, and the amplitude of flux variability. In particular, the variable EoS exhibits systematically larger temporal variability than the constant-$\gamma$ models.

These results demonstrate that adopting a physically self-consistent description of the EoS is essential for realistic modeling of accretion-flow thermodynamics and horizon-scale images like Event Horizon Telescope (EHT) observations.
\end{abstract}

\keywords{Black holes (162) --- Accretion (14) --- Magnetohydrodynamics (1964) --- Radiative transfer (1335) --- Plasma astrophysics (1261)}

\section{Introduction} 
\label{sec:intro}

The supermassive black holes at the centers of M87 and Sgr A* provide unique laboratories for studying strong gravity. Their horizon-scale images offer powerful constraints on black hole mass and spin while enabling stringent tests of general relativity \cite[GR,][]{einstein1915}. The Event Horizon Telescope (EHT) collaboration has achieved groundbreaking very long baseline interferometry (VLBI) observations at a wavelength of 1.3~mm, utilizing a baseline approximating the Earth's diameter through a network of radio telescopes \citep{EHT_M87_PaperI}. These observations have unveiled an asymmetric ring morphology of the central compact radio sources of M87 and Sgr A* \citep{EHT_M87_PaperI,EHT_M87_PaperII,Fromm-etal2022}. The results of both black holes align with the Kerr black hole (BH) model predicted by GR, reinforcing its validity \citep{EHT_M87_PaperV}.

Numerical investigations are indispensable in probing the physics of BH via shadow imaging \citep{EHT_M87_PaperIII,EHT_M87_PaperIV}. We employ general relativistic magnetohydrodynamic (GRMHD) simulations alongside general relativistic radiative transfer (GRRT) calculations to construct these shadow images. Several models and their corresponding computational codes have been developed to generate synthetic images, which are subsequently compared with observational data to find out the most accurate physical models \citep[e.g.,][]{RN21,Cruz-Osorio2022,RN14,Moscibrodzka-etal2016,Chael-etal2019}.

The shadow images are significantly influenced by electron synchrotron radiation, and the EHT collaboration assumes a Maxwellian thermal distribution for the electron distribution function (eDF) at the frequency of 230~GHz. Consequently, electron temperature becomes a critical parameter throughout the simulation process \citep{RN10}. In previous studies, the equation of state (EoS) for an ideal gas was extensively utilized to describe the plasma conditions \citep{Mizuno-etal2021}. The adiabatic index $\gamma$ was typically considered a constant, with $\gamma=4/3$ representing a relativistic (high-energy) gas and $\gamma=5/3$ corresponding to a non-relativistic (low-energy) gas. However, spatial and temporal variations in the gas state within the region of interest can lead to discrepancies in the temperature estimations. Therefore, a variable adiabatic EoS is introduced, adjusting according to gas temperature \citep{Ryu-Chattopadhyay2006,Mignone-McKinney2007,RN1}.

Ample numbers of semi-analytical studies have examined the role of plasma thermodynamics in accretion flows, showing that the choice of adiabatic index and the use of variable EoS can strongly influence transonic structure and shock formation in the accretion flow around black holes \citep[e.g.,][]{Chattopadhyay-Ryu2009,Vyas-etal2015,Dihingia-etal2018PhRvD,Dihingia-etal2019,Dihingia-etal2020}. Recently, two-temperature GRMHD simulation work by \cite{Sadowski-etal2017} demonstrated that evolving separate ion and electron thermodynamics, together with radiative processes, can significantly alter the temperature structure and thickness of radiatively inefficient accretion flows (RIAF), especially at intermediate accretion rates. More recently, \cite{Gammie_2025} showed that even single-fluid descriptions are sensitive to plasma microphysics, indicating that collisionless accretion flows favor an effective adiabatic index slightly below $\gamma=5/3$ and that the constant-$\gamma$ approximation can misrepresent trans-relativistic thermodynamics near the event horizon. Building on these developments, \cite{salas_2025} used two-temperature magnetically arrested disk (MAD) simulations of Sgr~A* to demonstrate that evolving the electron entropy and radiative cooling substantially reduces variability in millimeter wavelengths, which gives better agreement with observations than traditional single-temperature models.

We follow a two-temperature framework, which is widely considered essential to study RIAF and near-horizon radiative features of M87 and Sgr~A* \cite[e.g.,][]{Ressler-etal2015,RN19,Mizuno-etal2021,Dihingia-etal2023,RN30}. Additionally, when considering eDF, beyond thermal distributions, non-thermal effects such as magnetic reconnection and turbulent dissipation may drive electrons toward a power-law distribution, especially in the high-energy tail \citep{RN14}. We apply both the thermal Maxwell-J\"{u}ttner distribution and $\kappa$ distribution at two observation frequencies, 86~GHz and 230~GHz, for comparison \citep{RN21,Cruz-Osorio2022,Fromm-etal2022}.

In this work, we present a systematic GRMHD–GRRT study of how plasma microphysics, specifically the choice of EoSs and electron-heating prescription, shapes the thermodynamical closure, the subsequent dynamical evolution, and observable signatures of black hole accretion flows. By comparing traditional constant-$\gamma$ EoSs ($\gamma = 4/3, 5/3$) with the variable EoS, we show the thermodynamic response of the flow in both accretion flow states: MAD and standard and normal evolution (SANE) regime. These microphysical differences propagate into measurable variations in shadow morphology, image size, sub-millimeter flux, spectral slopes, and variability. 
Our results demonstrate that the choice of thermodynamic closure introduces systematic differences in the flow dynamics and horizon-scale emission. These differences may arise from both the direct thermodynamic response and indirect dynamical effects mediated by EoS-dependent disk geometry and initial conditions, as also reported by \cite{Mignone-McKinney2007,White_2020,B_gu__2023}.

This paper focuses on M87* as the primary target, and the rest of the paper is structured as follows: Section~\ref{sec:intro} provides an overview of the background. In Section~\ref{numerical}, we provide details of the numerical modelling. Sections~\ref{results01}-\ref{results03} present a comparative analysis of different models. Finally, in Section~\ref{sec:conclusion}, we summarize and discuss their impacts in astrophysical contexts.

\section{Numerical setup} \label{numerical}
\subsection{GRMHD simulations}

To compare the effects of different EoSs, we employ the Black Hole Accretion Code ({\tt BHAC}) \citep{RN40,RN29} to perform three-dimensional (3D) general relativistic magnetohydrodynamic (GRMHD) simulations. For this study, we utilized the two-temperature module developed by \citet{Mizuno-etal2021} in the {\tt BHAC} code. The initial conditions for the simulations also closely follow \cite{Mizuno-etal2021}, where they used a rotationally supported hydrostatic Fishbone-Moncrief (FM) torus \citep{Fishbone-Moncrief1976,Uniyal-etal2024}. 

The initial FM torus for each model is constructed with its corresponding adiabatic index ($\gamma = 4/3$, $\gamma = 5/3$, or $\gamma = 13/9$ for the variable case)
and spin parameter $a=0.94$. Although the models share the same prescribed inner edge and density maximum radius, the EoS-dependent enthalpy-to-density mapping produces measurable differences in the initial density distribution, disk thickness, radial extent, and total torus mass, which are shown in detail in Appendix~\ref{appendixD}.
The total mass in the high-density region differs by only $\approx 10\%$ among the EoS models. These small initial differences may still influence the subsequent accretion, magnetic flux transport, heating, and resultant emission.

For the torus under the SANE case, its weaker magnetic-flux accumulation makes it rather sensitive to the initial magnetic-flux reservoir and geometry. But for the MAD case, the simulation starts from a much larger torus and quickly becomes regulated by the saturated magnetic flux near the black hole. 
The larger MAD torus extends well beyond the region primarily responsible for the horizon-scale emission, which may reduce sensitivity to its outer boundary. Nevertheless, differences in the initial geometry and magnetic-flux supply cannot be completely excluded.

The torus is defined by the inner edge at $r_{\rm in}=6\,r_g$ and the radius for density maximum at $r_{\rm max}=12\,r_g$ for the SANE case, where $r_g=GM/c^2$ is the gravitational radius. For the MAD case, we use $r_{\rm in}=20\,r_g$ and $r_{\rm max}=40\,r_g$ for making a larger initial torus.

Throughout the study, lengths are scaled with $r_g$ and, similarly, time is scaled with $t_g=r_g/c$, where $M$, $G$, and $c$ are the mass of the central BH, the universal gravitational constant, and the speed of light, respectively. Along with the FM torus, we supply an initial vector potential to introduce a magnetic field, which is given by
\begin{align} 
A_\phi &\propto \max\left(q-0.2,0\right),\\
q &=
    \begin{cases}
      \frac{\rho} {\rho_{\rm max}}, \   &\mathrm{for\ \ SANE} \\
      \frac{\rho} {\rho_{\rm max}}\left(\frac{r}{r_{\rm in}}\right)^3 \sin^3 \theta \ \exp \left(-\frac{r}{400}\right),  &\mathrm{for\ \ MAD.}\\
    \end{cases}  
    \label{eq:Aphi}
\end{align}

For both models, we initialize a single poloidal magnetic loop confined within the torus (Eq.~\ref{eq:Aphi}). In MAD models, we employ a radially concentrated vector potential designed to largely fill the magnetic field in the torus.
The initial magnetic field strength is set by specifying the minimum initial plasma-$\beta$, 
which indicates the ratio of gas and magnetic pressure ($\beta=2p/b^2$). We apply $\beta_{min}=100$, which is the minimum value of the ratio with co-located values.
%
%

After building up the initial torus, in the consecutive evolution we implemented the variable adiabatic index EoS, where the adiabatic index depends on the local temperature $(\Theta=p/\rho)$. In general, in most of the previous studies, a simple ideal EoS with a constant adiabatic index is extensively used, so-called constant-$\gamma$ or ideal EoS \citep{Rezzolla-Zanotti2013}. The value of the constant adiabatic index $\gamma$ depends on the consideration of plasma gas temperature and composition. For a non-relativistic temperature ($\Theta\ll1$), it is $\gamma$=5/3. On the other hand, if the gas reaches a relativistic temperature ($\Theta\gg1$), it becomes $\gamma$=4/3. However, in the realistic situation, the gas can be trans-relativistic, so the effective adiabatic index need not be fixed at either limiting value, and the adiabatic index could be $4/3\leq\gamma\leq5/3$. To solve this problem, \cite{Synge1957} introduced an EoS based on kinetic theory, which gives the enthalpy ($h$) of the gas as:
\begin{equation}
    h_{Synge}=\frac{K_3(1/\Theta)}{K_2(1/\Theta)},
    \label{eossynge}
\end{equation}

where $K_2$ and $K_3$ are the modified Bessel functions of the second kind. Direct evaluation of Bessel functions in the Synge EoS is computationally expensive and time-consuming for numerical simulations. The TM EoS provides a computationally efficient approximation to the relativistic Synge gas, enabling a smooth transition between non-relativistic ($\gamma \approx 5/3$) and relativistic ($\gamma\approx4/3$) regimes without explicitly evaluating Bessel functions \citep{Mignone-McKinney2007}, which is given by
\begin{equation}
    h=\frac{5}{2}\Theta+\sqrt{\frac{9}{4}\Theta^2+1}.
    \label{eostm}
\end{equation}
For this EoS, the adiabatic index of the gas can be expressed as
\begin{equation}
    \gamma_{\rm eff}=\frac{1}{6}(8-3\Theta_{gas}+\sqrt{9\Theta_{gas}^2+4}).
    \label{gamma}
\end{equation}
Similarly, the adiabatic indices for electrons and ions can be expressed as 
\begin{align}
    &\gamma_{\rm e,eff}=\frac{1}{6}(8-3\Theta_{e}+\sqrt{9\Theta_{e}^2+4}), {~~\rm and}
    \label{gammae} \\
    &\gamma_{\rm i,eff}=\frac{1}{6}(8-3\Theta_{i}+\sqrt{9\Theta_{i}^2+4}).
    \label{gammai}
\end{align}

\begin{figure}
\centering
		\includegraphics[width=\linewidth]{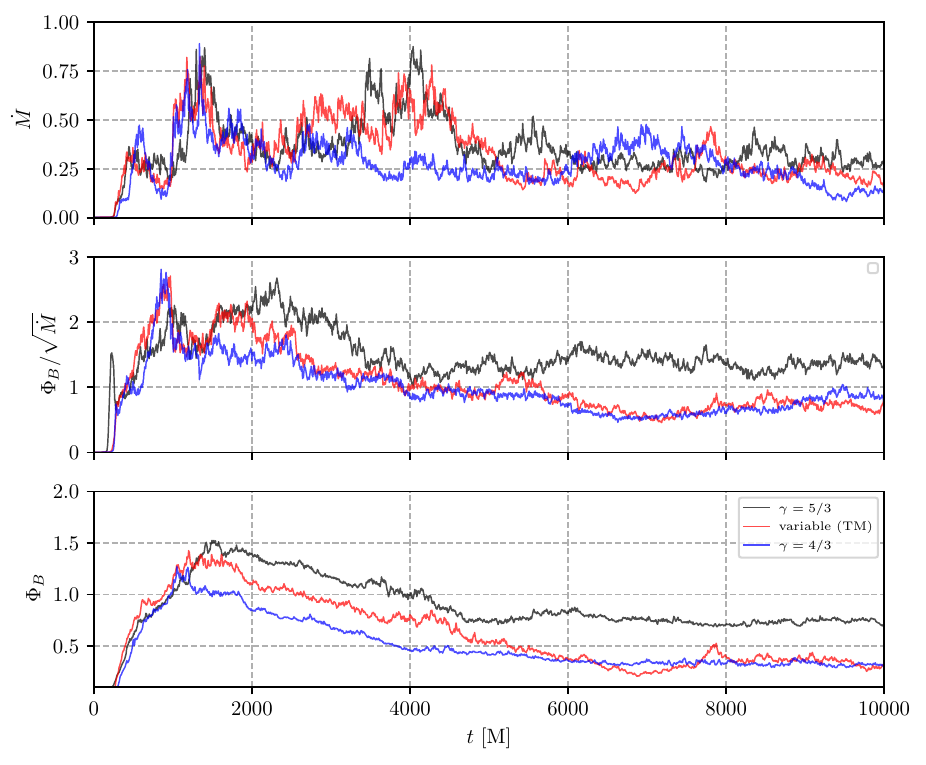}
    \caption{Time evolution of mass accretion rate, normalized magnetic flux rate, and magnetic flux rate for SANE cases with different EoS; the constant-$\gamma$ EoS with $\gamma=4/3$ (blue), with $\gamma=5/3$ (black), and the variable EoS (red).}
    \label{3d1}
\end{figure}

\begin{figure}
\centering
		\includegraphics[width=\linewidth]{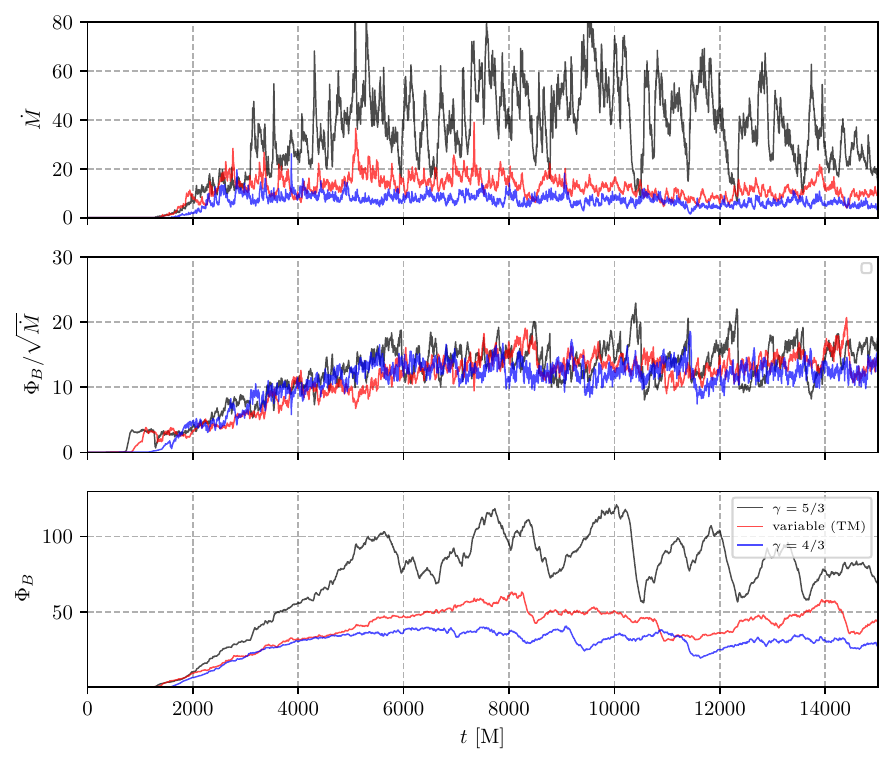}
    \caption{Same as Fig.~\ref{3d1} but shown for MAD cases. }
    \label{3d2}
\end{figure}

Here, the respective temperatures are defined by corresponding individual pressures and densities, i.e., $\Theta_k=p_k/\rho_k$ ($k=g, i, e$), where the subscripts $g$, $i$, and $e$ stand for the gas, ions, and electrons, respectively. With this, we solve the GRMHD equations in a simulation domain from inside the horizon up to $r=2500\,r_g$ with an effective resolution of $384\times192\times192$ with 3 static mesh refinement layers.

In our simulation, the static mesh refinement is prescribed geometrically following \cite{Mizuno-etal2021}. Blocks at larger radius \(r>100\,r_g\), as well as blocks close to the polar axis near the horizon \(r<8\,r_g\) with \(\theta<\pi/10\) or \(\theta>\pi-\pi/10\), are forced to coarsen to a lower resolution. 
The remaining regions, including the torus and near-equatorial flow, are refined to provide higher spatial resolution to capture the disk dynamics and near-horizon accretion flow well.

The simulations are evolved up to $t=15000\,t_g$. We devised three simulation models in the SANE as well as MAD regime (a total of six cases). Two cases correspond to the extreme limits for the adiabatic index with ideal EoS, i.e., $\gamma=4/3$ and $\gamma=5/3$, and another case of TM EoS. For the ideal constant-$\gamma$ EoSs, we fix the electron and ion adiabatic indices to be $\gamma_e=4/3$ and $\gamma_i=5/3$, respectively.

In addition to the single-fluid GRMHD equations, we couple with the electron thermodynamics using an electron entropy equation to obtain the electron temperature locally, following the approach used in previous two-temperature GRMHD simulations \citep[e.g.,][]{Ressler-etal2015,Sadowski-etal2017, Mizuno-etal2021}. 
In this work, we consider two electron-heating prescriptions based on turbulent heating and magnetic-reconnection heating, following the subgrid models of \citet{Howes2010}, \citet{Rowan-etal2017}, and \citet{Kawazura-etal2020}. For each of the six GRMHD setups, corresponding to three EoS prescriptions in both SANE and MAD states, we apply both electron-heating prescriptions. We note that we ignore the radiative cooling and Coulomb coupling effects, which would change electron thermodynamics \citep[e.g.,][]{Dihingia-etal2023,Zhang-etal2026}


\subsection{GRRT calculations}

Subsequently, we utilize our GRMHD simulations to perform general relativistic radiative transfer (GRRT) calculations to visualize the impacts of EoSs on the radiative properties. For this study, we consider the target source to be M87*, which is a widely studied supermassive black hole. Based on current observations, the accretion flow around M87* can be modeled as MAD \citep{EHT_M87_PaperI,EHT_M87_PaperV}. The adopted parameters are: BH mass $M=6.5\times10^9M_\odot$, spin $a=0.94$, distance $D=16.8\times10^3$ kpc, and inclination angle $i=163^\circ$. We employ the GRRT code {\tt BHOSS} \citep{Younsi_2012,Younsi_2020,Younsi_2023} to solve the geodesic equations in curved space-time and then solve the radiative transfer equation along the null geodesics to produce synthetic images. We use the synchrotron process to calculate emissions from the near-horizon region. Accordingly, we use the field of view (FoV) of $\pm 130\,r_g$ in both directions, which is approximately equal to $\pm 500\,\mu as$. Moreover, we set the targeted flux density to $\sim 1.0$~Jy at $230$~GHz, while the resolution is set to be 500$\times$500 pixels. 

This targeted flux density value lies between the compact flux density at 230~GHz inferred by \citep{EHT_M87_PaperIV}, \(F_{\rm cpct}=0.66^{+0.16}_{-0.10}\,\mathrm{Jy}\), and the larger arcsecond-scale core flux density, \(F_{\rm tot}\simeq1.2\,\mathrm{Jy}\). Our FoV is larger than the horizon scale of EHT observations but smaller than the arcsecond scale. We expect a non-negligible fraction
of the 230~GHz flux to come from outside the compact EHT-scale emission region. Therefore, we choose this target flux value.
We conduct the investigation within the simulation time $t=13000-15000\,t_g$ with steps of $10\,t_g$.

In this study, we examine the impact of electrons on radiative properties, including variability, image morphology, and extended emission at $86\,$GHz and $230\,$GHz. Accordingly, we consider the thermal electron distribution function (eDF) and non-thermal $\kappa$ eDF, which are given by
\begin{align}
    &\frac{\mathrm{d} n_e}{\mathrm{d} \tilde{\gamma}_e}=\frac{n_e}{\Theta_e}\frac{\tilde{\gamma}_e\sqrt{\tilde{\gamma}_e^2-1}}{K_2(1/\Theta_e)}\exp\left(-\frac{\tilde{\gamma}_e}{\Theta_e} \right), ~~{\rm and}
    \label{eDF-thermal} \\
    &\frac{\mathrm{d} n_e}{\mathrm{d} \tilde{\gamma}_e}=N\tilde{\gamma}_e \sqrt{\tilde{\gamma}_e^2-1}\left(1+\frac{\tilde{\gamma}_e-1}{\kappa w}\right)^{-(\kappa+1)}, 
    \label{eDF-kappa}
\end{align}
respectively. Here $n_e$, $\tilde{\gamma}_e$, $\Theta_e$, $N$ are the number density, the Lorentz factor, the dimensionless temperature of the electrons, and the normalization factor, respectively. In Eq.~\ref{eDF-kappa}, $w$ depends on $\kappa$ and the electron temperature as

\begin{equation}
    w =
        \begin{cases}
            \frac{\kappa-3}{\kappa}\Theta_e , \ & \kappa \ge3 \\
            \Theta_e, \ & \kappa < 3 \\
        \end{cases}  
\label{kappa_w}\
\end{equation}

Finally, $\kappa$ is expressed as \citep[e.g.,][]{Meringolo-etal2023}, and we also set a threshold for $\sigma_{cut}=1$ and maximum value of $\kappa=7.5$. 
During GRRT post-processing, cells with $\sigma>\sigma_{cut}$ are excluded from the radiative-transfer calculation by neglecting their emissivity and absorptivity, because their thermodynamic variables may be affected by numerical floor prescriptions. This cutoff does not modify the underlying GRMHD variables.
\begin{equation}
    \kappa =3.8+0.2\sigma^{-1/2}+1.6\sigma^{-0.6}\tanh(2.25\sigma^{1/3}\beta)  \label{kappa}\,.
\end{equation}

\begin{figure*}[h]
\centering
\includegraphics[width=0.8\linewidth]{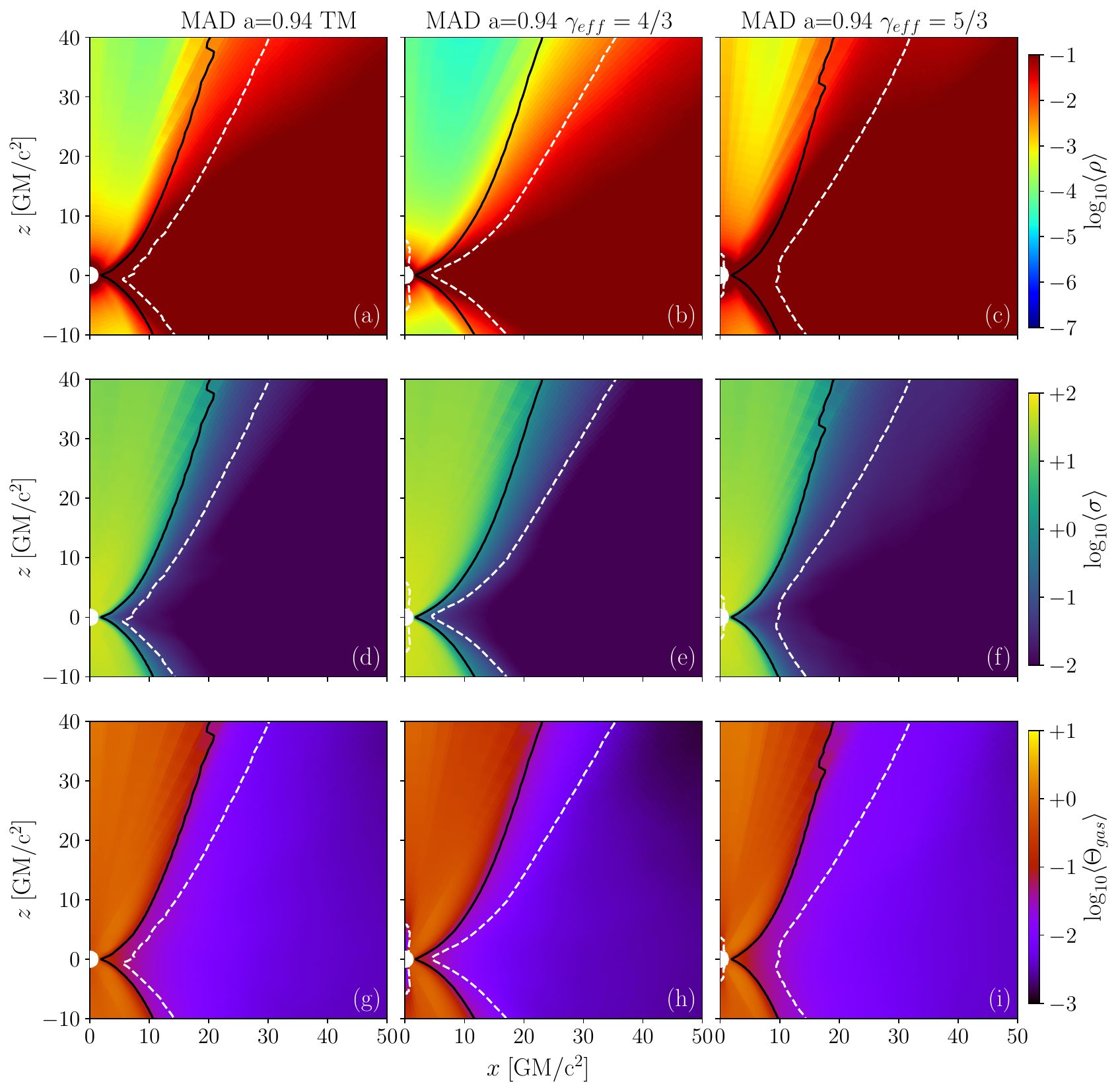}
\caption{The logarithms of time and azimuthally averaged density (upper panels), magnetization (middle panels), and gas temperature (lower panels) distributions on the poloidal plane for the MAD models for different EoSs (left: variable EoS, middle: constant $\gamma$ EoS with $\gamma=4/3$, right: with $\gamma=5/3$). black-solid and white-dotted lines correspond to magnetization $\sigma=1$ and Bernoulli parameter $-hu_t=1.02$, respectively. For the time average, we choose $t=13,000-15,000\,M$.
}
\label{fig:MADflow}
\end{figure*}

\section{Fluid Dynamics}
\label{results01}

\begin{figure*}[h]
\centering
\includegraphics[width=0.85\linewidth]{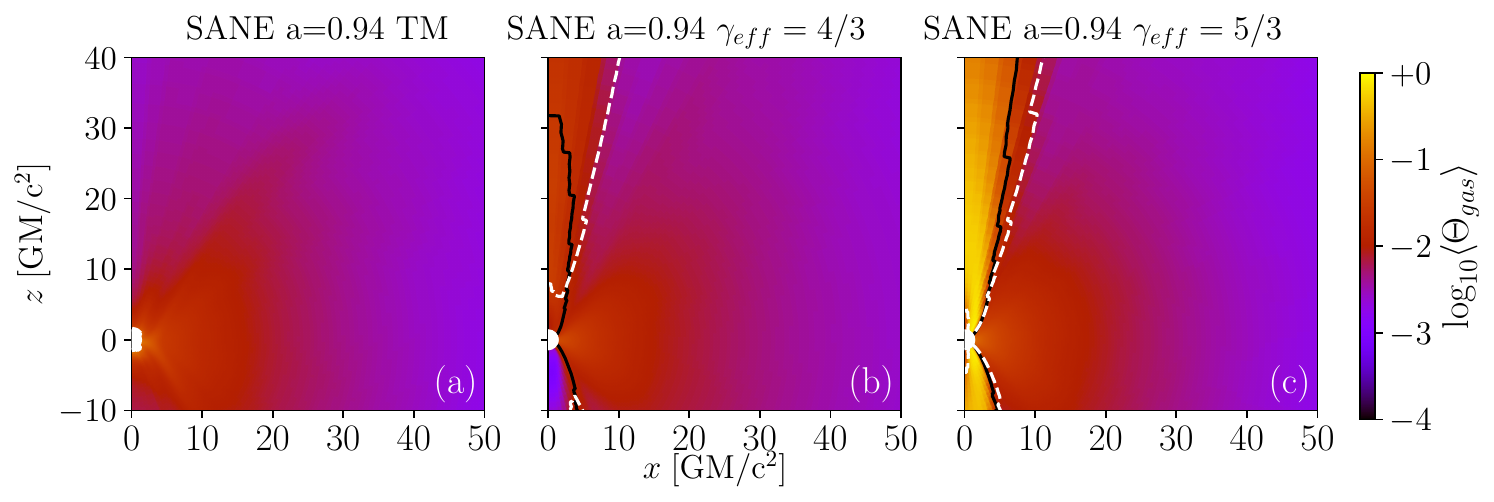}
\caption{Same as Fig.~\ref{fig:MADflow}, but for SANE models, and only gas temperature is shown.}
\label{fig:SANEflow}
\end{figure*} 

Before investigating electron thermodynamics and radiative properties, we first analyze the large-scale fluid properties of our simulations. The temporal behavior of mass accretion rate and magnetic flux establishes the dynamical state of SANE and MAD flows. 

Accordingly,
in Figs.~\ref{3d1} and \ref{3d2}, we show the time evolution of the accretion rate ($\dot{M}$, upper panel), normalized magnetic flux ($\Phi_B/\sqrt{\dot{M}}$, middle panel), and magnetic flux ($\Phi_B$, lower panel) calculated at the horizon for SANE and MAD cases, respectively. In different panels, blue, red, and black lines correspond to the profiles calculated for ideal EoS with $\gamma=4/3$, TM EoS, and ideal EoS with $\gamma=5/3$, respectively. Note that these quantities do not depend on the electron heating prescriptions. Therefore, we do not compare them here.

The time evolution of these quantities shows similar qualitative behavior in both the SANE and MAD models. That is, they start to grow initially and reach a quasi-steady state, then remain there till the end of the simulations. 
We also observe a characteristic MAD-state balance between the gas and magnetic pressures, providing lateral evidence that the flow has entered the magnetically arrested regime.
Additional evidence that the inner MAD flow has reached an approximately quasi-steady state is presented in Appendix ~\ref{appendixC}.

One noticeable difference in these diagnostics
between SANE and MAD models is that the time required to reach quasi-steady state for MAD models is longer than that of the SANE models because of the larger inner radius of the MAD initial setup.

Additionally, we note that irrespective of SANE or MAD models, $\gamma=5/3$ cases show higher accretion rates, normalized magnetic flux, and magnetic flux at the horizon than those of the other models, i.e., $\gamma=4/3$ and TM (variable) EoS. For SANE models, the normalized fluxes remain within $\Phi_B/\sqrt{\dot{M}}\sim1$, confirming them to be within the expected limits \citep{RN40}. Similarly, for MAD models, the normalized fluxes remain within $\Phi_B/\sqrt{\dot{M}}\sim12-15$, which is the expected range of MAD models \citep[e.g.,][]{Mizuno-etal2021}.

Besides the time evolution profiles of mass accretion rate and magnetic flux rate under SANE and MAD states, we cut the 3D simulation result into 2D slices to show the distribution of various physical quantities in Fig.~\ref{fig:MADflow} for MAD models. The first, second, and third columns correspond to quantities calculated for TM EoS, constant-$\gamma$ EoSs with $\gamma=4/3$, and with $\gamma=5/3$, respectively. On the other hand, the first, second, and third rows correspond to the logarithmic values of time and azimuthal-averaged density $(\rho)$, magnetization ($\sigma=b^2/\rho$), and gas temperature ($\Theta_{\rm gas}=p/\rho$), where $b^2=b^\mu b_\mu$ and $b^\mu$ is the magnetic 4-vector. The black solid and dashed lines correspond to the boundary of magnetization $\sigma=1$ and Bernoulli parameter $-hu_t=1.02$, respectively. The region with $\sigma>1$ and $-hu_t>1$ roughly corresponds to the jet region \citep[e.g.,][]{Nathanail-etal2020,Dihingia-etal2021}.

The funnel region ($\sigma \gtrsim 1$) exhibits a clear dependence on the adopted thermodynamical closure and the resulting flow evolution. Among the three cases, the constant-$\gamma$ EoS with $\gamma = 4/3$ provides the widest funnel region, while the constant-$\gamma$ EoS with $\gamma = 5/3$ produces the least extended one. The TM EoS lies in between these two extremes. A similar trend is observed in the opening angle, as illustrated in panels (a)–(c) of Fig.~\ref{fig:MADflow}. Analyzing the magnetization distribution (Fig.~\ref{fig:MADflow}(d)-(f)), we find that the funnel region is most strongly magnetized for the $\gamma = 4/3$ case, weakest for the $\gamma = 5/3$ case, and intermediate for the TM EoS. In contrast, the temperature distribution ($\Theta_{\rm gas} = p/\rho$, Fig.~\ref{fig:MADflow}(g)-(i)) shows the opposite behavior: the gas temperature is highest for $\gamma = 5/3$, lowest for $\gamma = 4/3$, and again intermediate for the TM EoS.

Additionally, we have performed a similar study for SANE models. We observe quite similar qualitative behaviors for density, magnetization (with lower maximum values), and gas temperature. To avoid repetition, in Fig.~\ref{fig:SANEflow}, we only show the gas temperature $\Theta_{\rm gas}$ for TM EoS, constant $\gamma$ EoS with $\gamma=4/3$, and with $\gamma=5/3$ in panels (a), (b), and (c), respectively. Thus, we see that the gas temperature is maximum, intermediate, and minimum for $\gamma=5/3$, TM, and $\gamma=4/3$, respectively. 

These EoS-dependent differences in the disk and funnel structures are broadly consistent with previous GRMHD studies. \citet{Mignone-McKinney2007} found that different adiabatic-index prescriptions can modify the turbulence, magnetic-field structure, and outflow geometry. Similarly, \cite{B_gu__2023} reported a weak dependence of the jet geometry on the adiabatic index and noted that changing $\gamma$ also modifies the initial torus thickness and mass. Our results likewise indicate that the EoS influences both the initial torus geometry (Appendix~\ref{appendixD}) and the subsequent dynamical evolution.

In summary, 
under both SANE and MAD states, the $\gamma=5/3$ case yields higher temperatures, whereas the $\gamma=4/3$ case yields lower temperatures, relative to the TM EoS case. 
The fixed enthalpy-temperature relation of the constant-$\gamma$ EoSs contributes directly to these temperature differences. However, EoS-dependent changes in disk geometry and vertically structured transport may also affect the evolved thermodynamic state especially in SANE case.

In this sense, the constant-$\gamma$ models sample systematic differences associated with the choice of thermodynamical closure.

On the other hand, variable EoSs, such as TM EoS, are consistent to be utilized in studying relativistic accretion flow around black holes. In the next sections, we will study the impacts of these overestimations/underestimations on electron temperature and radiative properties. 

\begin{figure*}[ht]
\centering
\includegraphics[width=0.8\linewidth]{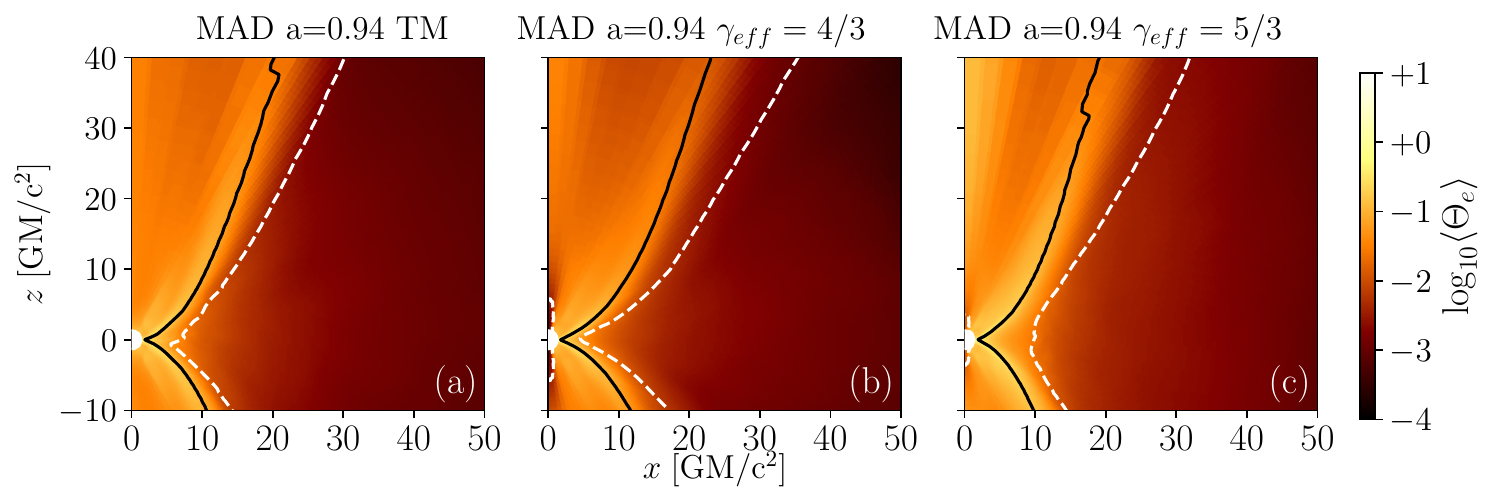}
\caption{Same as Fig.~\ref{fig:SANEflow}, but shown electron temperature $\Theta_{\rm e}$ for MAD models.}
\label{fig:MADTe}
\end{figure*}

\begin{figure*}[ht]
\centering
\includegraphics[width=0.8\linewidth]{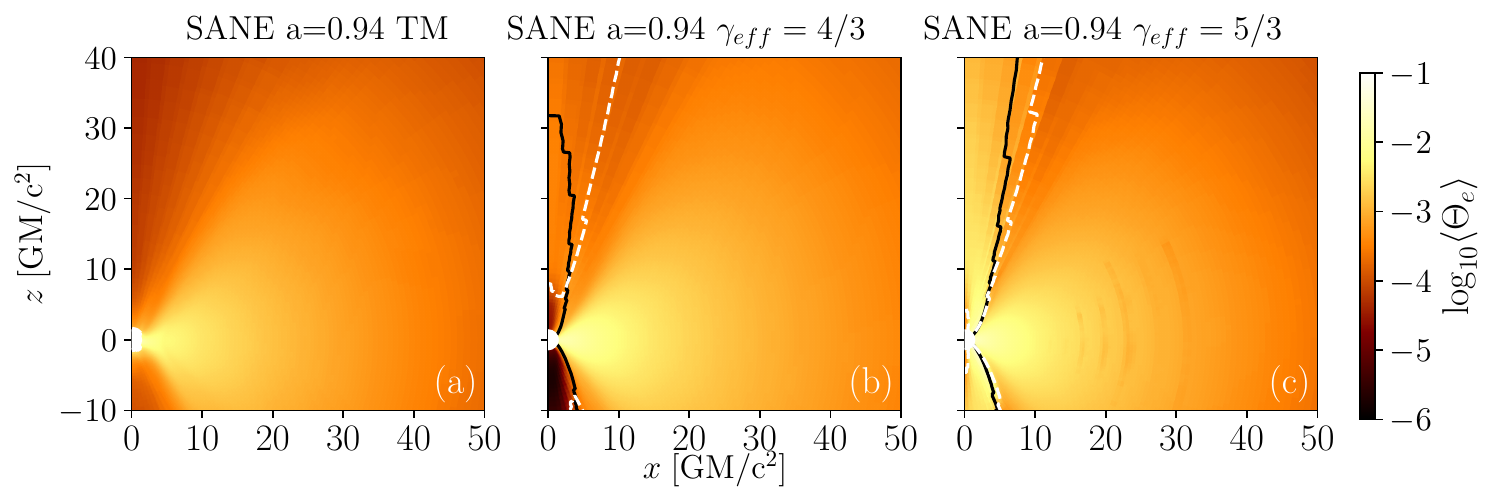}
\caption{Same as Fig.~\ref{fig:MADTe}, but shown for SANE models.}
\label{fig:SANETe}
\end{figure*}
\section{Properties of electrons}
\label{result02}

In the previous section, we observed that the flow temperature depends on the choices of EoS. Since radiative properties are highly sensitive to the electron temperature ($\Theta_{\rm e}$), understanding how different EoSs modify $\Theta_{\rm e}$ is essential for interpreting the resulting images. We therefore, in this section, examine the behavior of electron temperatures in both SANE and MAD cases.

In Figs.~\ref{fig:MADTe} and \ref{fig:SANETe}, we display the dimensionless electron temperature for MAD and SANE cases, respectively. The arrangements of panels are the same as in Fig.~\ref{fig:SANEflow}. Here, we focus on turbulent heating cases for illustration.
Note that, for panels \ref{fig:MADTe}(b) and \ref{fig:MADTe}(c), the adiabatic index of electrons for both cases is $\gamma_e=4/3$. 
Since electrons in such a scenario are expected to be transrelativistic to ultrarelativistic, we decide not to study them by varying $\gamma_e$ to different values.

\begin{figure*}
    \centering
    \includegraphics[width=0.95\linewidth]{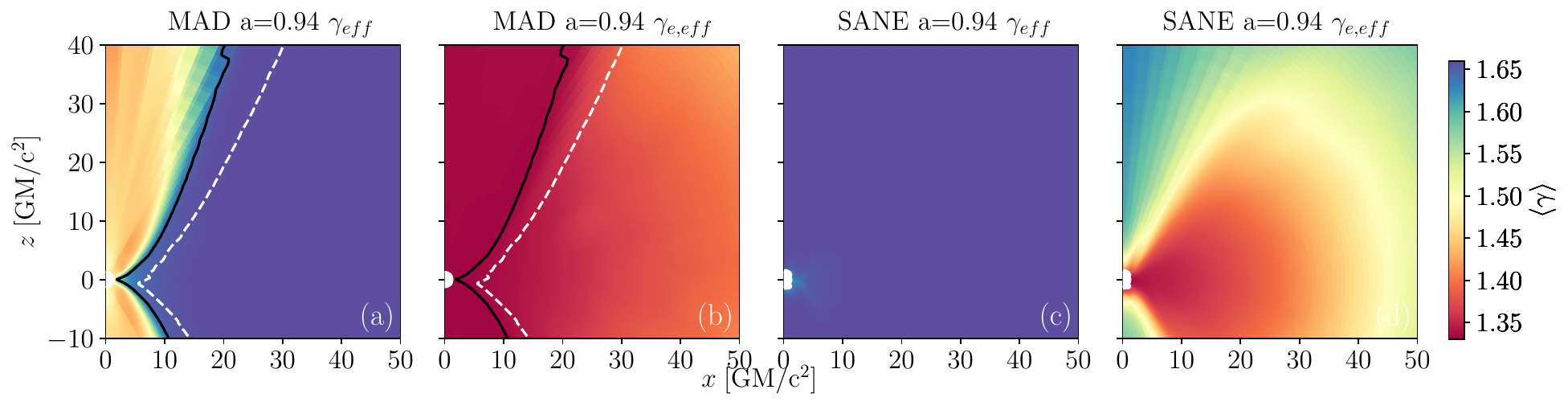}
    \caption{Time and azimuthal-averaged distribution of the effective $\gamma$ value of plasma gas and electrons under MAD (left panels (a and b)) and SANE state (right panels (c and d))}
    \label{variousgamma}
\end{figure*}

For MAD cases, the electrons are heated up to very high temperatures due to the turbulent heating. Accordingly, even for TM EoS $\gamma_{\rm e,eff}$ is approximate to 4/3 throughout the simulation domain (also see Fig.~\ref{variousgamma}), particularly in and around the funnel region. Therefore, these regions have quite similar electron temperatures in all three panels of Fig.~\ref{fig:MADTe}. However, in the region $\sigma<1$ (high-density region), we observe an overestimated electron temperature, which can influence the extended shadow images of black holes (e.g., $86$ GHz and $230$ GHz).

On the contrary, for the SANE models, the tori are less strongly magnetized, so electron heating processes are less efficient, as seen in Fig.~\ref{fig:SANETe}. Accordingly, we observe drastically different electron temperatures for TM (panel \ref{fig:SANETe}a), $\gamma_{\rm eff}=4/3$ (panel \ref{fig:SANETe}b), and $\gamma_{\rm eff}=5/3$ (panel \ref{fig:SANETe}c) EoS cases. As a result, $\gamma$ values for TM EoS are much higher than $\gamma_{\rm eff}=4/3$. This leads to a lower electron temperature for TM EoS. However, for $\gamma_{\rm eff}=5/3$, we observe a much hotter funnel region. In summary, the electron temperature in SANE cases is therefore particularly sensitive to the adopted thermodynamic closure, and the TM EoS provides a self-consistent treatment of the trans-relativistic regime.

Finally, to understand the realistic ranges of the adiabatic index for the fluid and electrons, in Fig.~\ref{variousgamma}, we show $\gamma_{\rm eff}$ and $\gamma_{\rm e,eff}$ for  SANE (right panels, \ref{variousgamma}c,d) and MAD models (left panels, \ref{variousgamma}a,b). These figures suggest that the adiabatic index for the fluid in the high-density region is always $\gamma_{\rm eff}\sim5/3\sim1.66$. Whereas, in the funnel region, the adiabatic index is slightly lower; for the MAD model, it is of the order of $\gamma_{\rm eff}\sim3/2\sim1.5$. Consistent with the expectation, the electron adiabatic index for SANE cases is much higher than $\gamma_{\rm e,eff}>4/3$ in most of the simulation domain and approaches 4/3 only very close to the black hole. In MAD cases, by contrast, the value of the adiabatic index is $\gamma_{\rm e,eff}\sim4/3$ in most of the simulation domain up to $10$s of gravitational radii. 

As a result, considering $\gamma=5/3$ and $\gamma_{\rm e}=4/3$ being a good approximation for MAD models, other fixed-\(\gamma\) choices may therefore introduce larger systematic differences in the inferred electron thermodynamics.
SANE models need TM EoS for consistent fluid and electron temperature estimation. 

\section{radiative properties}
\label{results03}

The shadow images are expected to depend on the electron distribution functions, temperature, and magnetic field. We now quantify how different EoSs and electron-heating prescriptions affect the images at 230 and 86 GHz using the GRRT {\tt code BHOSS} \citep{Younsi_2012,Younsi_2020,Younsi_2023}. This allows us to directly connect microphysics in the GRMHD simulations to differences in spectral energy distribution (SED), image morphology, light-curve variability, etc.

In this section, we present the time-averaged shadow images, light curves, total flux time variability, and spectral energy distribution curves for comparative analysis.
Electron temperatures are obtained from the GRMHD simulations utilizing two different heating prescriptions based on turbulence and magnetic reconnection.

\begin{figure*}[h]
    \centering\includegraphics[width=0.8
    \linewidth]{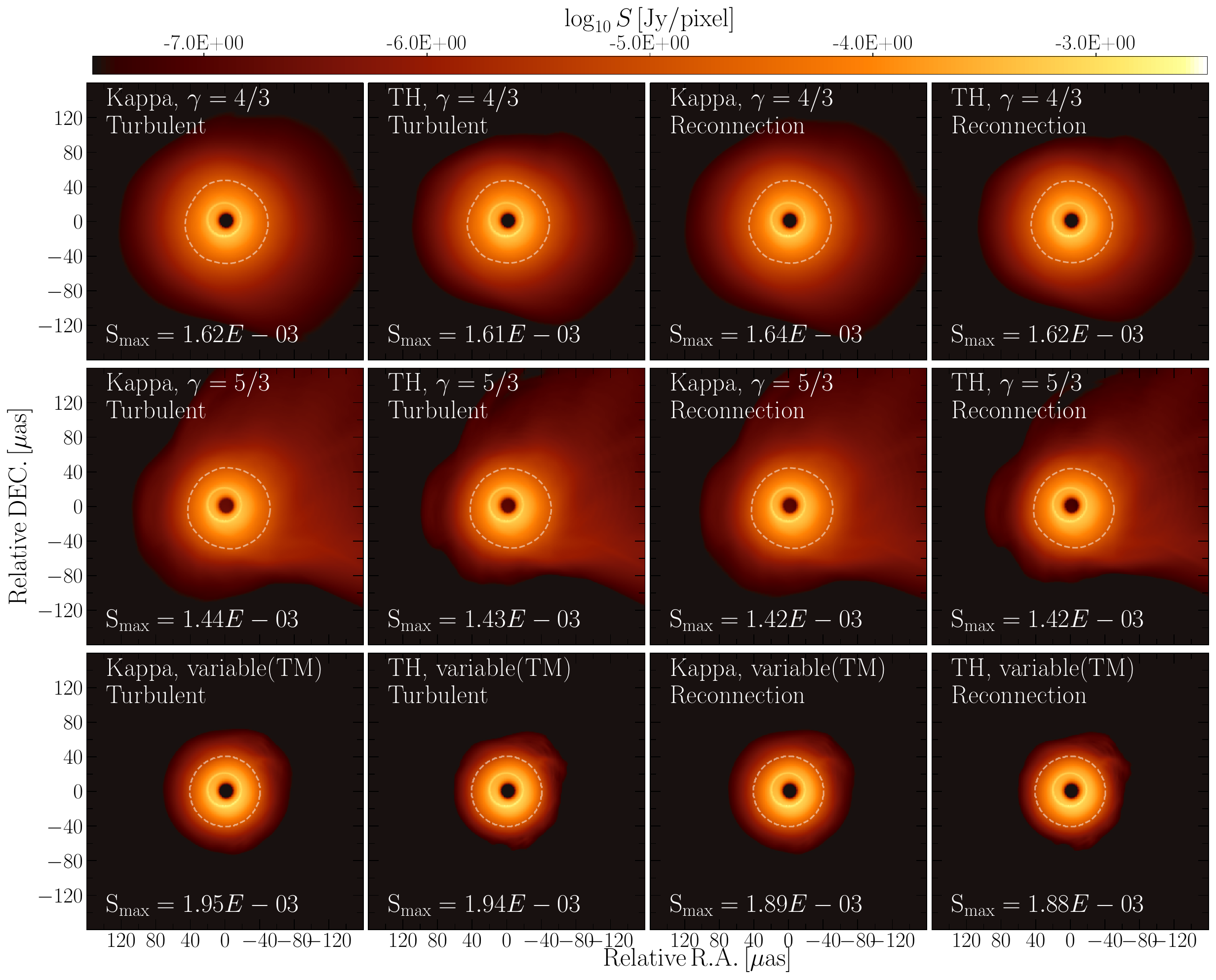}
  \caption{Time-averaged intensity images at 230~GHz for the MAD state under different EoSs, two-temperature models, and two electron distributions (thermal and $\kappa$) with a logarithmic scale. The white dashed contour in the panels corresponds to $0.01$ of the maximum intensity.}
  \label{mad-230}
\end{figure*}

\begin{figure*}[ht]
    \centering
\includegraphics[width=0.8\linewidth]{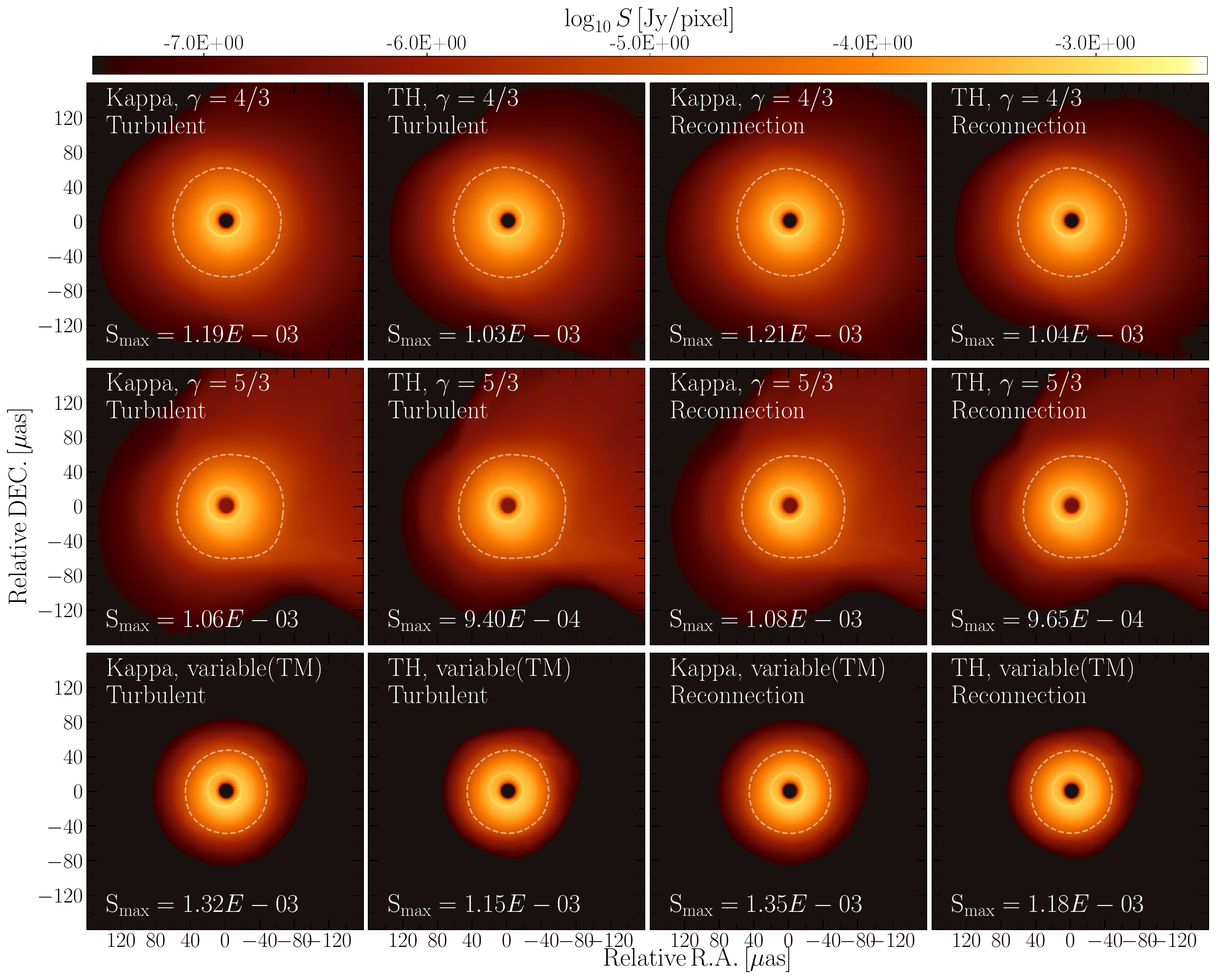}
  \caption{Same as Fig.~\ref{mad-230} but shown for 86~GHz. 
  }
  \label{mad-86}
\end{figure*}

Figures~\ref{mad-230} and \ref{mad-86} show time-averaged images for MAD models at 230~GHz and 86~GHz with different EoSs and electron-heating prescriptions, respectively. All models reproduce the well-known and expected bright ring morphology, indicating that the spacetime geometry primarily determines the overall size and shape of the image. Nonetheless, distinct and methodical variations manifest in the brightness distribution, ring thickness, and emission extent, illustrating the impact of plasma thermodynamics.
The TM EoS produces a smaller and more concentrated emission ring, with weaker diffuse emission at larger radii.
This is also supported by the one-dimensional intensity profiles discussed in detail in Appendix~\ref{appendixB}, which demonstrate that the maximum intensity is higher in the TM case than in constant-$\gamma$ EoS cases.

By contrast, the constant-$\gamma$ EoS models produce broader structures. For example, the $\gamma = 4/3$ case shows a wider funnel-dominated emission, while the $\gamma = 5/3$ case increases emission from the dense disk, making the ring thicker and less clearly defined. This behavior can be explained by looking at synchrotron emissivity, which is sensitive to the temperature profile of the electrons and the magnetic field. The TM EoS allows for a smooth and self-consistent transition between relativistic and non-relativistic regimes, which focuses emission in the inner regions. On the other hand, constant-$\gamma$ cases cause systematic differences in temperature and, as a result, in emissivity. We observe similar noticeable differences in the images for SANE models, particularly at 86 GHz (see Appendix~\ref{appendixA}). 

The image differences are more pronounced at 86 GHz because lower-frequency emission arises from larger radii and more optically thick regions. As a result, the 86 GHz images are more sensitive to the thermodynamic structures of the flow. Because of this, constant-$\gamma$ models show more extended emission than the TM case, especially in areas where temperature differences are biggest. This suggests that the choice of EoS can affect long-wavelength observables more strongly than higher-frequency, near-horizon emission. 

The time-averaged flux at 86~GHz is higher than the time-averaged flux at 230~GHz in all models. This behavior is expected because the 86~GHz emission samples a larger, more optically thick volume of the flow, while the 230~GHz emission comes from the optically thin part of the flow closer to the center. The size of this frequency dependence changes relying on the choice of EoS. 

We also compare thermal and $\kappa$ eDFs together with different electron-heating prescriptions. At 230~GHz, the differences between thermal and $\kappa$ distributions are modest, as the emission is dominated by near-horizon regions where electrons are already relativistic.
At 86~GHz, the high-energy tail of the \(\kappa\) distribution can enhance emission from larger radii, producing smoother and more extended intensity profiles \cite[e.g.,][]{RN6}.
Nevertheless, Figs.~\ref{mad-230} and \ref{mad-86} suggest that such eDF-dependent differences are comparatively small. Similarly, differences between turbulent and magnetic reconnection heating models are comparatively subdominant in shaping the image morphology. 
The image differences are more strongly associated with the overall EoS-dependent flow structure than with the other two eDFs or electron-heating prescriptions. This structure reflects both the thermodynamic closure and its geometry-mediated dynamical consequences.

\begin{figure*}[ht]
    \centering
\includegraphics[width=0.80\linewidth]{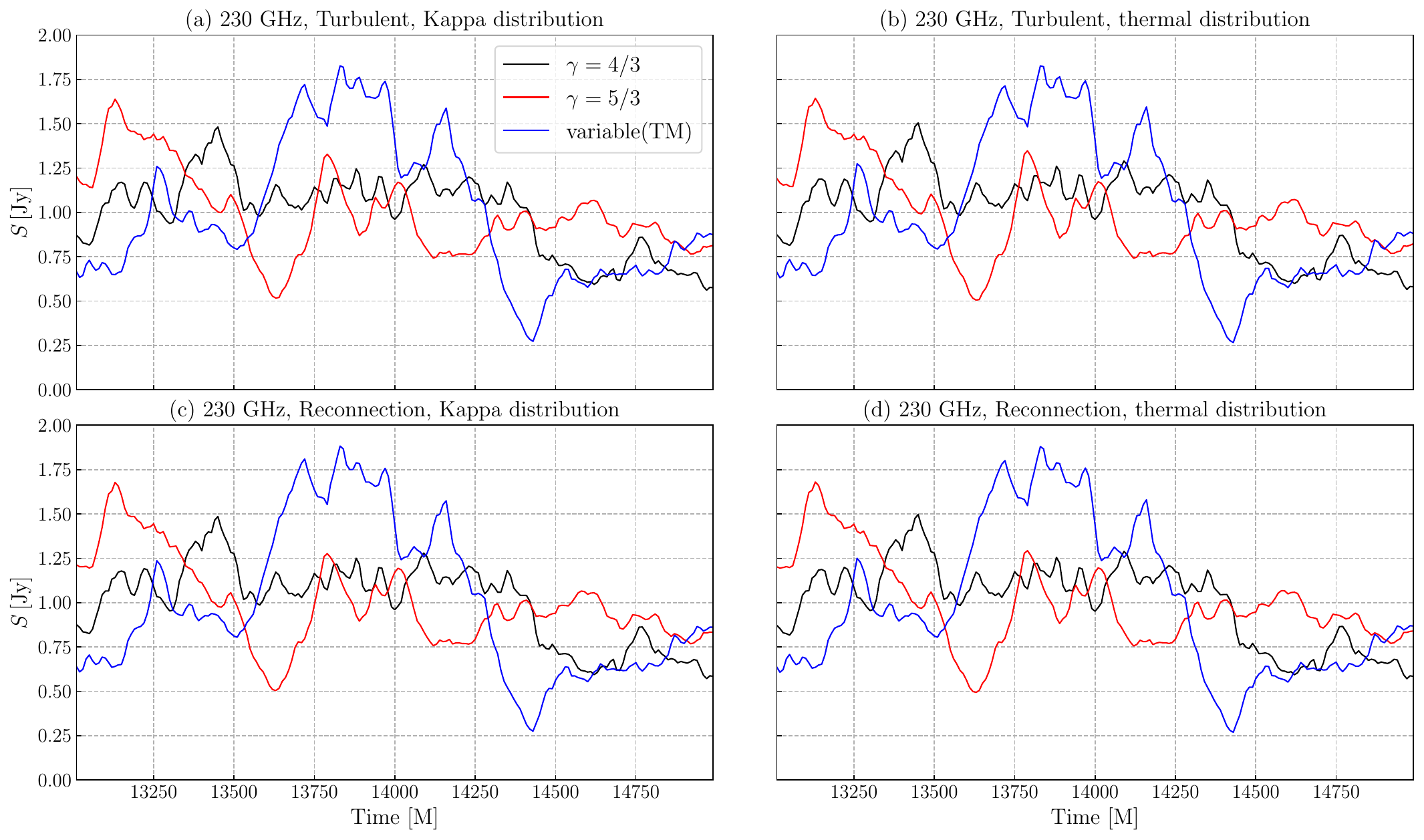}
\includegraphics[width=0.80\linewidth]{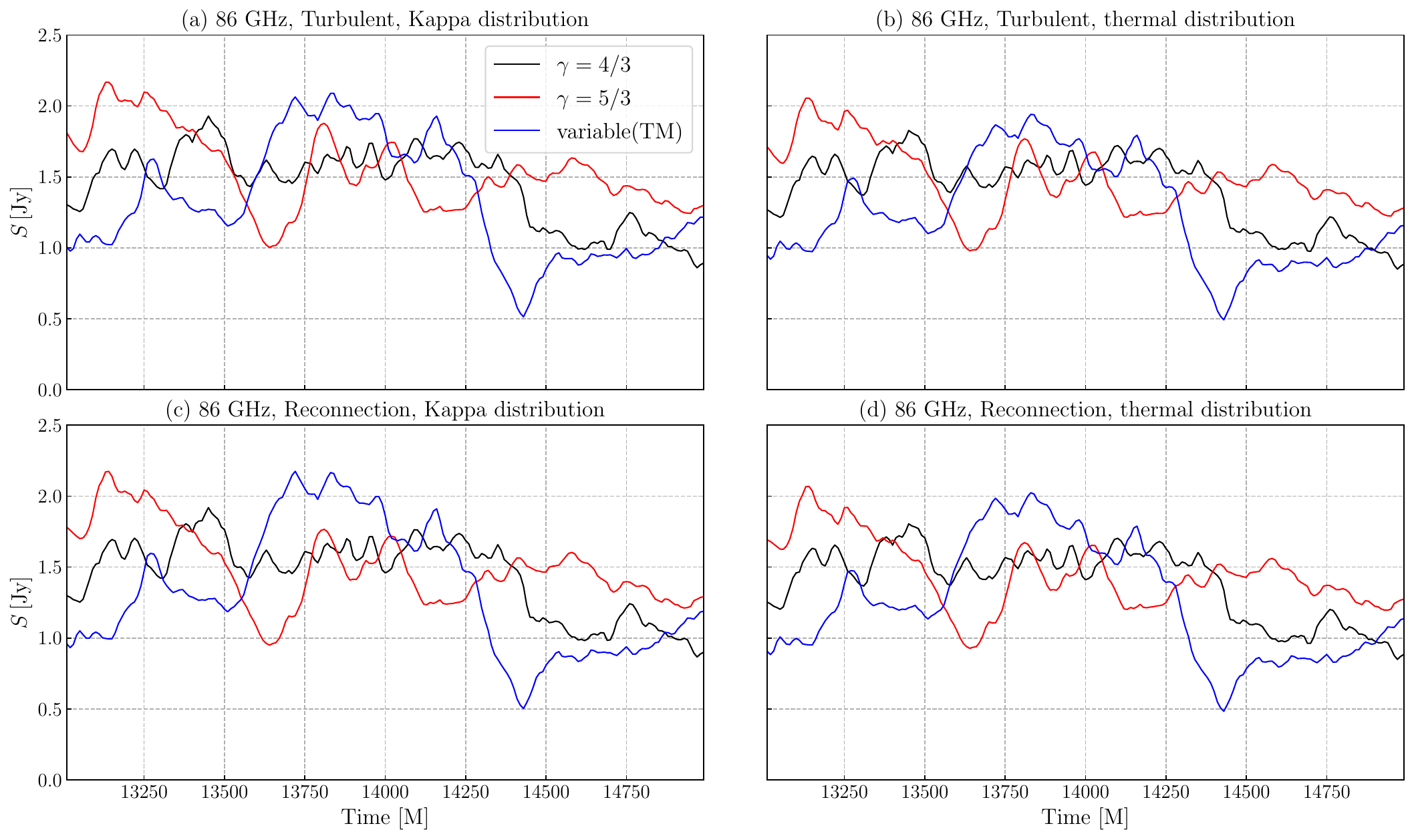}
  \caption{Light curve at 230~GHz (upper) and 86~GHz (lower) under MAD state, calculated separately with a two-temperature model with turbulent and magnetic reconnection heating prescriptions, combined with three various EoSs during the time period of 
  $13\,000$ -$15\,000 M$.}
  \label{lightcurve}
\end{figure*}

\begin{figure*}[ht]
    \centering
	\includegraphics[width=0.80\linewidth]{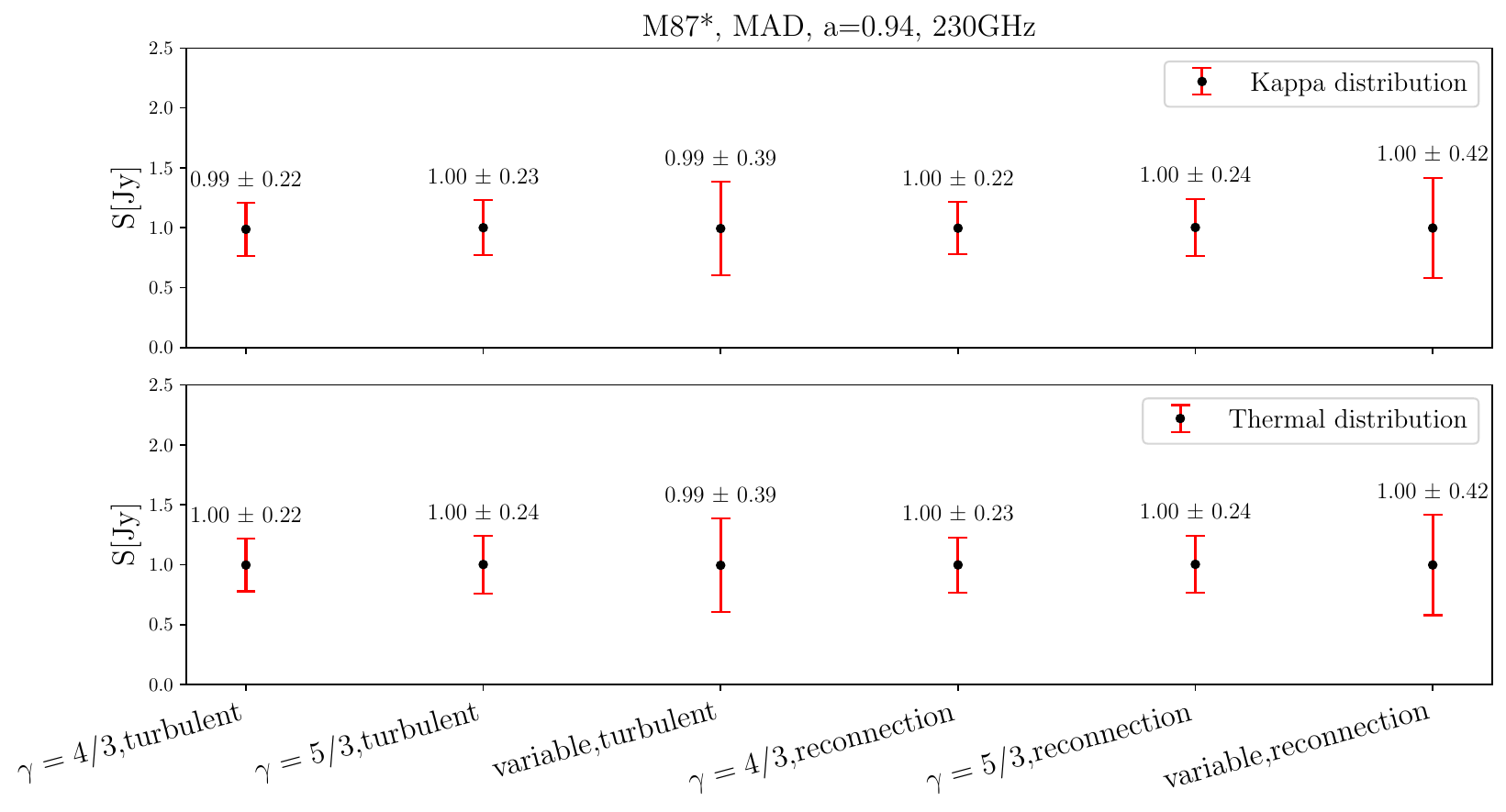}
    \includegraphics[width=0.80\linewidth]{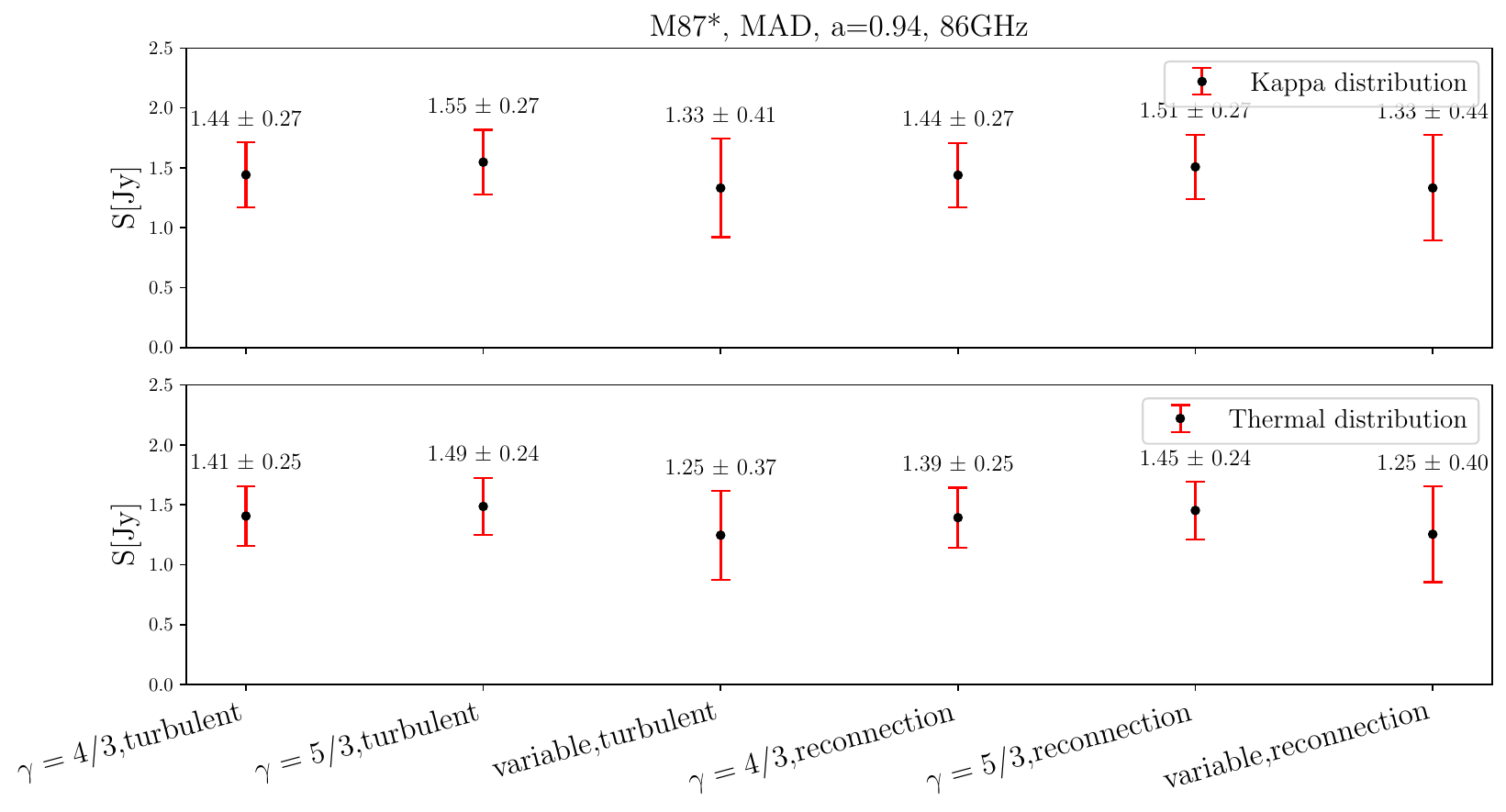}
  \caption{Total flux time variation at 230~GHz (upper) and 86~GHz (lower) under MAD state with $i=163^\circ$, calculated separately with two-temperature models with turbulent and magnetic reconnection heating in different EoSs.The upper panel adopted $\kappa$ eDF while the lower one applied the thermal eDF. All dots in this figure represent the average value of total intensity in each case, and the error bars signify the standard deviation relative to the average values.}
  \label{variabilityfig}
\end{figure*}

\begin{figure*}
    \centering
    \includegraphics[width=0.46\linewidth]{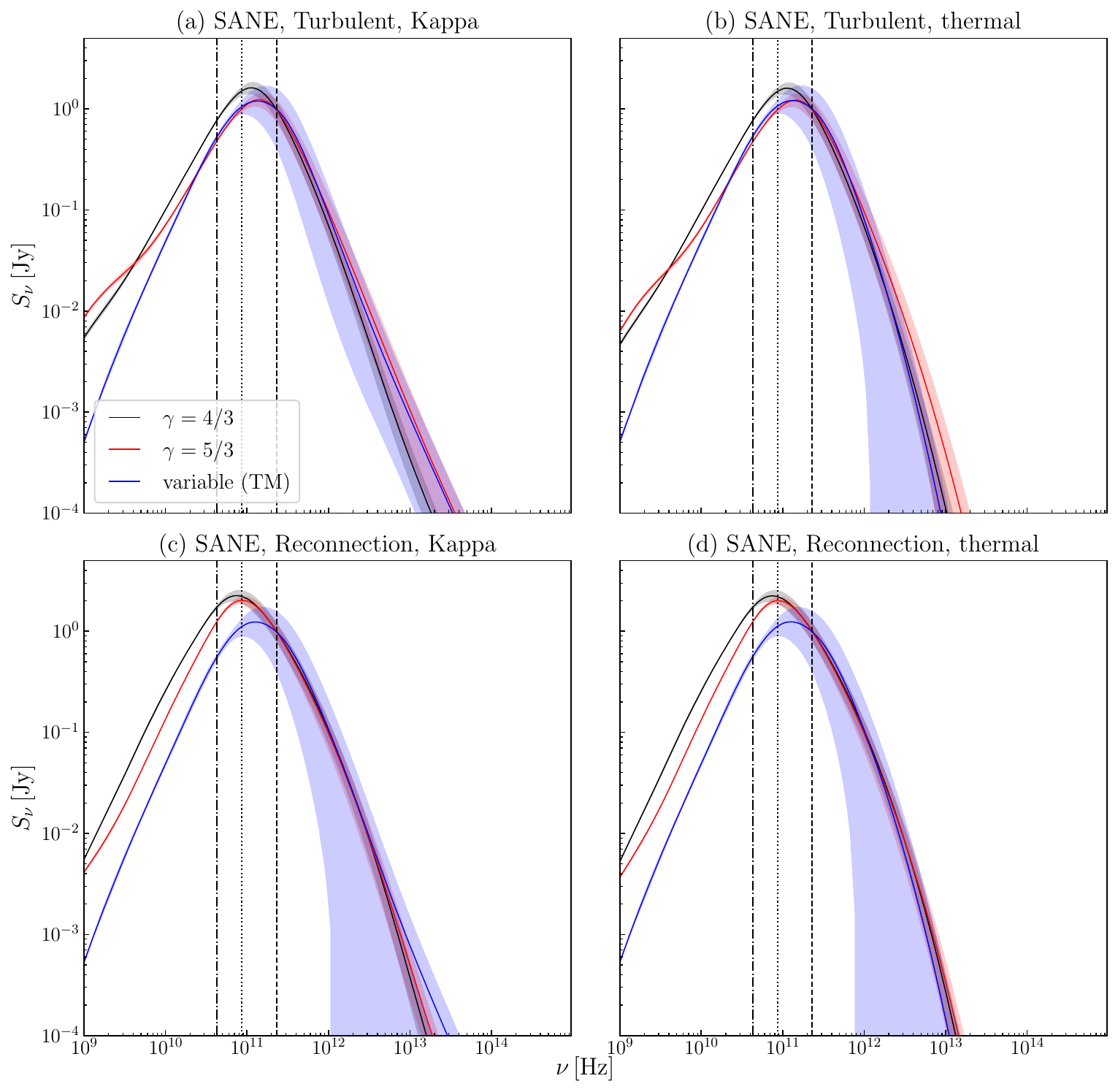}
    \includegraphics[width=0.46\linewidth]{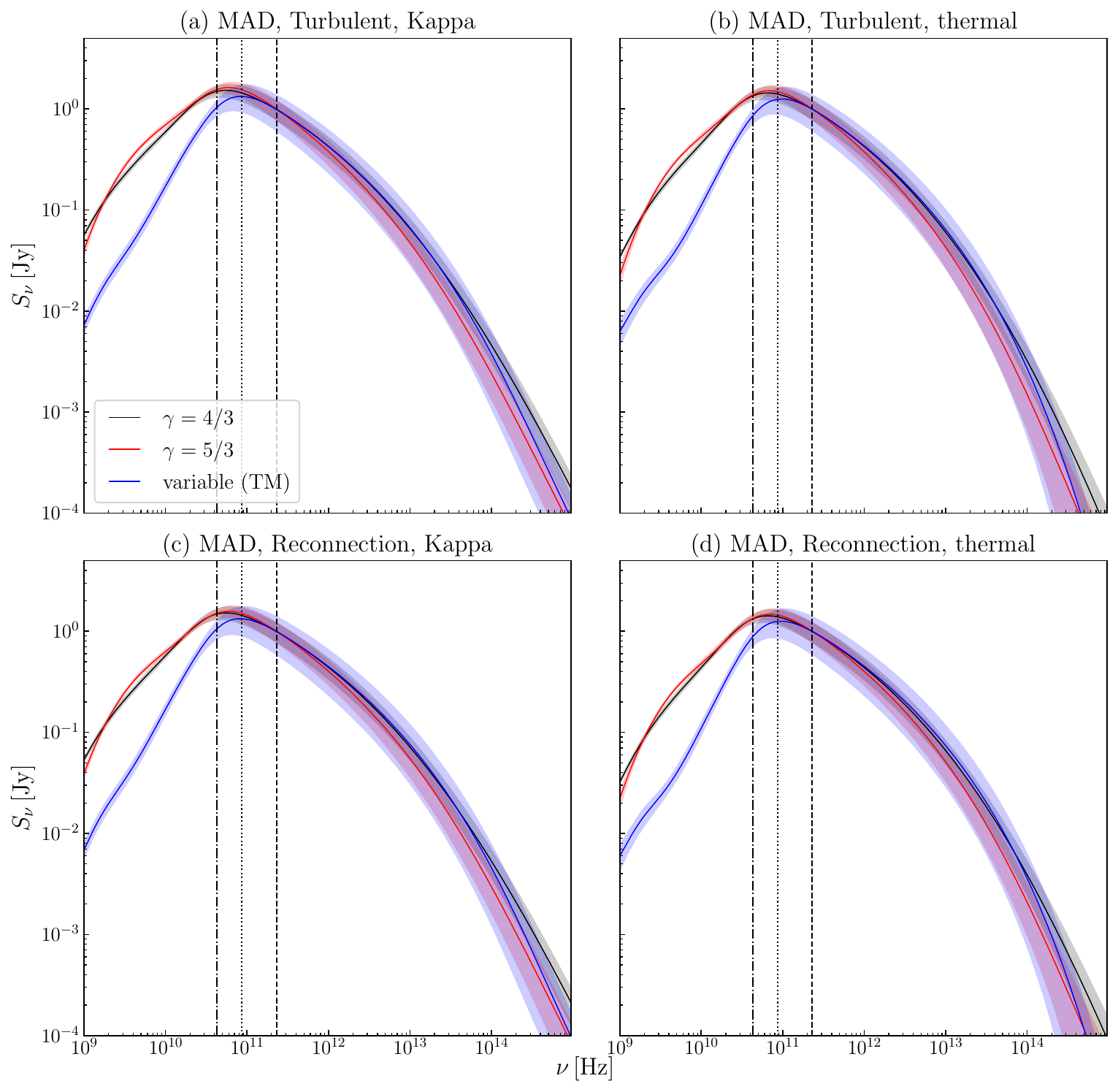}
    \caption{Spectral energy distribution with different EoSs under SANE (left) and MAD (right) states with different electron distributions: $\kappa$ eDF ({\it a,c}) and thermal distribution ({\it b,d}). The black, red, and blue curves represent three EoSs: $\gamma=4/3$, $\gamma=5/3$, and the variable TM EoS, respectively. The solid curve is the average value, while the shaded region around the curve represents the standard deviation relative to the average value. The dashed and dotted vertical lines correspond to 43, 86, and 230~GHz.}
    \label{spectrum}
\end{figure*}

Fig.~\ref{lightcurve} shows light curves of the emission for different simulation models, and Fig.~\ref{variabilityfig} presents the corresponding variability amplitudes. The TM case shows larger flux variability than the constant-$\gamma$ cases at both 230~GHz and 86~GHz over the analyzed time interval. Figure~\ref{variabilityfig} quantifies this trend through the flux standard deviation: the TM model has the value $\approx 0.39$--$0.42$ at 230~GHz and $\approx 0.30$--$0.33$ at 86~GHz, compared to $\approx 0.22$--$0.24$ and $\approx 0.16$--$0.19$, respectively, for the constant-$\gamma$ models --- nearly a factor of two larger, robustly across both electron-heating prescriptions and both eDFs. Identifying the physical driver of this enhanced variability would require a dedicated time-lag analysis of the horizon magnetic flux, accretion rate, and thermodynamic averages, which we defer to future work. This trend is measured over the finite time interval analyzed here, and longer simulations are required to determine whether it persists as a converged long-term property.

Fig.~\ref{spectrum} shows the spectral energy distributions (SEDs), which further illustrate the impact of plasma microphysics on the radiative properties.
The SEDs are computed from the multi-frequency output of {\tt BHOSS}. For each GRMHD snapshot and model, we perform ray-tracing at 128 logarithmically spaced frequencies between \(10^9\) and \(10^{15}\) Hz. At each frequency, {\tt BHOSS} returns the corresponding specific-intensity map on the image plane. We integrate the intensity over all image pixels to obtain the total flux density \(F_\nu\), and then construct the SED from these frequency-dependent fluxes.

All models show the typical synchrotron spectrum, but the normalization and spectral slope change depending on the EoS. Quantitatively, at 86~GHz the time-averaged flux density of the TM model ($1.25$--$1.33$~Jy, depending on the heating prescription and eDF) is $\approx 10$--$15\%$ lower than that of the constant-$\gamma$ models ($1.39$--$1.55$~Jy). Around the SED peak ($\nu \approx 10^{11}$~Hz) the three EoS models agree to within $\approx 20$--$25\%$, and at high frequencies ($\nu \gtrsim 10^{12}$~Hz) to within a factor of $\approx 1.5$, largely within the time-variability bands. At low frequencies ($\nu \lesssim 10^{10}$~Hz) the TM model lies a factor of $\approx 2$--$3$ below the constant-$\gamma$ models, which are nearly indistinguishable from each other.

We interpret these differences as systematic theoretical uncertainties associated with thermodynamic response and geometry-mediated effects, which should be considered when modelling multi-frequency observations. The differences in the average spectra are more visible in the lower frequencies, rather than in the higher frequencies. Observationally, our results could be utilized to investigate discrepancies in observations of long wavelengths. A systematic comparison of modeling and observation is needed for further comment on the applicability of constant-$\gamma$ accretion flow.

\section{Conclusion and discussion}\label{sec:conclusion}

Based on our GRMHD simulations with different EoSs and electron-heating prescriptions, coupled with GRRT calculations at 86 and 230 GHz for M87*, we summarize the main findings as follows:
\begin{itemize}
    \item[1] The two constant-$\gamma$ models produce systematic uncertainties and differences in gas and electron temperatures relative to the TM model. $\gamma = 5/3$ case predicts excessively high temperatures in dense regions, while $\gamma = 4/3$ case yields lower temperatures and a more extended magnetized funnel, whereas the TM EoS provides intermediate, physically consistent trans-relativistic behavior.
    
    \item[2] In MAD simulations, strong magnetization drives electrons to ultra-relativistic temperatures ($\Theta_e$) due to heating processes largely independent of the adopted EoS. In the SANE simulations, however, the weaker magnetic field makes the electron temperature substantially more sensitive to the adopted thermodynamical models. 
    Consequently, the $\gamma = 4/3, \gamma = 5/3$, and TM EoSs produce markedly different thermal structures in the SANE regime.
    
    \item[3] Compared with the constant-$\gamma$ model, the TM EoS produces a brighter and more concentrated ring structure with a less extended jet, demonstrating $\gamma$ value changes with the surrounding environment. These variations alter the conversion and retention of thermal energy within the accretion flow.

    \item[4] Time variability at both 230~GHz and 86~GHz is larger for the TM EoS than for the constant-$\gamma$ cases over the analyzed time interval (The standard deviation value $\approx 0.39$--$0.42$ vs.\ $0.22$--$0.24$ at 230~GHz; $\approx 0.30$--$0.33$ vs.\ $0.16$--$0.19$ at 86~GHz). Establishing the causal origin of this enhancement requires longer simulations and dedicated time-dependent diagnostics.
    
    \item[5] We observed considerable differences at the lower frequencies. Thus, our investigations demonstrate that the choice of constant-$\gamma$ EoS introduces systematic theoretical uncertainties --- up to a factor of $\approx 2$--$3$ in the low-frequency SED and a factor of $\approx 2$ in flux variability --- that must be accounted for when interpreting EHT observations.
\end{itemize}

Our results highlight the importance of adopting a physically motivated EoS for plasma thermodynamics when connecting GRMHD simulations to the EHT observations \citep{EHT_M87_PaperIV,EHT_M87_PaperV}. In particular, the sensitivity of emission features to the EoS implies that microphysical assumptions can influence how horizon-scale images and spectra are interpreted \citep{EHT_M87_PaperV,Fromm-etal2022}. Even though the overall shape of the shadow stays the same, differences in brightness distribution, variability magnitude, timescale, and extended emission show that similar observational signatures can come from different plasma conditions. This adds another level of degeneracy to modeling efforts \citep[e.g.,][]{Narayan-etal2012,RN6} and calls for caution when using current and future observations to constrain physical parameters \citep[e.g.,][]{Mitra_2022,salas_2025,Davelaar-etal2018}.

These results are particularly significant for the interpretation of multi-frequency observations of sources like M87* and Sgr~A*. The larger dependency on thermodynamics at lower-frequency spectra indicates that long-wavelength observations (e.g., 86~GHz and below) can effectively assess plasma conditions at larger radii. The increased variability magnitude linked to TM EoS indicates that time-domain observations might offer an alternative method to differentiate between various thermodynamical models (\cite{Vyas-etal2015,Ryu-Chattopadhyay2006}). 
Future efforts should combine multi-frequency imaging, time-domain observations, and improved theoretical modeling to break these degeneracies and better constrain the microphysics of black-hole accretion flows.

\begin{acknowledgments}
This research is supported by the National Key Research
and Development Program of China (grant no. 2023YFE0101200), the National Natural Science Foundation of China (grant nos.12273022, 12511540053, 123B1007, W2641030), the Shanghai Municipality Orientation Program of Basic Research for International Scientists (grant no.22JC1410600), and the Zhiyuan Future Scholar Program (grant no.ZIRC2023-03).
The simulations were performed on the Astro
cluster in the Tsung-Dao Lee Institute, $\pi$ 2.0, and Siyuan-1 cluster in the Center for High Performance Computing at Shanghai Jiao Tong University.
\end{acknowledgments}

\appendix
\section{Images for SANE models}
\label{appendixA}

\begin{figure*}[ht]
    \centering  
    \includegraphics[width=0.7\linewidth]{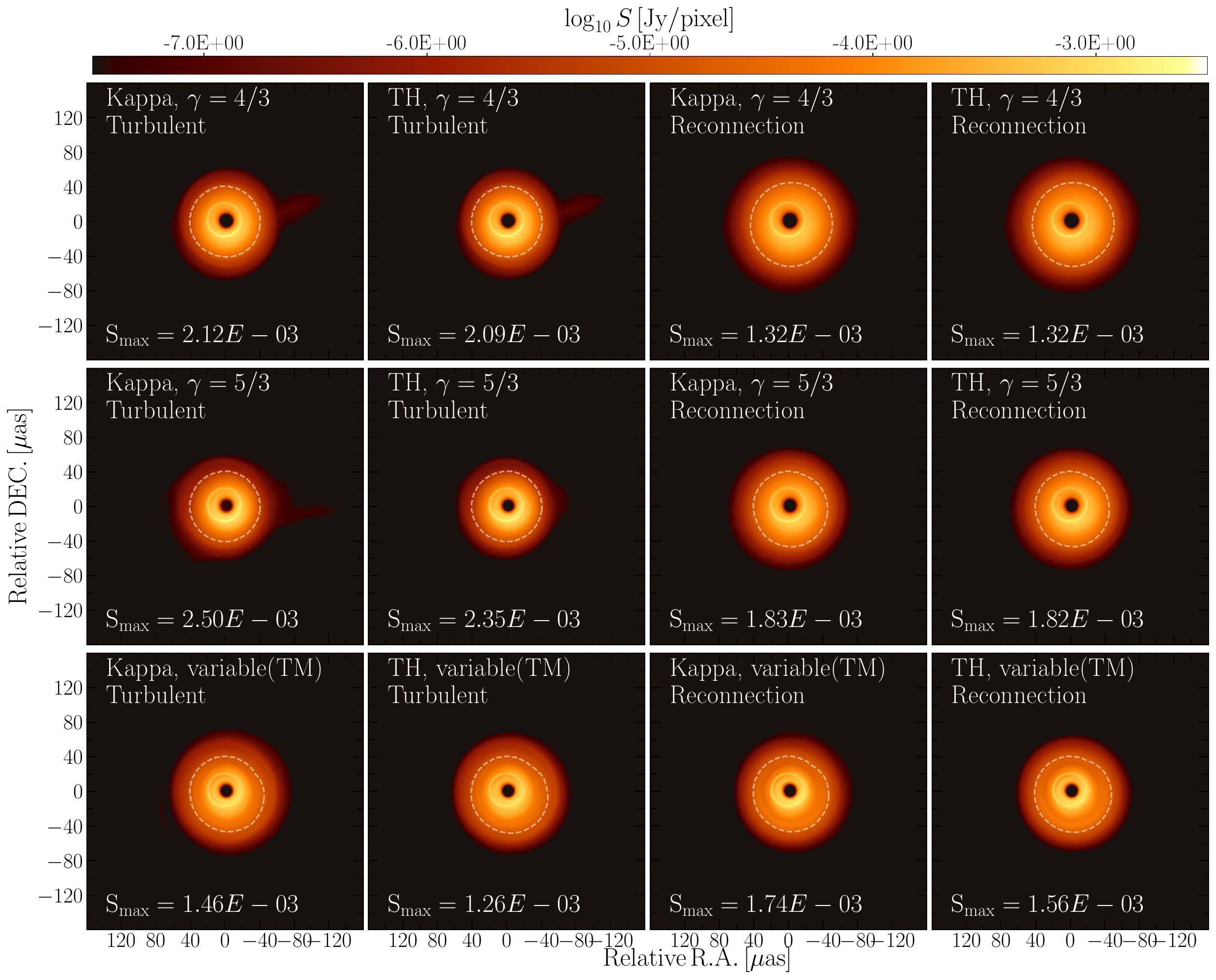}
  \caption{Same as Fig.~\ref{mad-230} but shown for SANE models.}
  \label{sane-230}
\end{figure*}

\begin{figure*}[ht]
    \centering
        \includegraphics[width=0.7\linewidth]{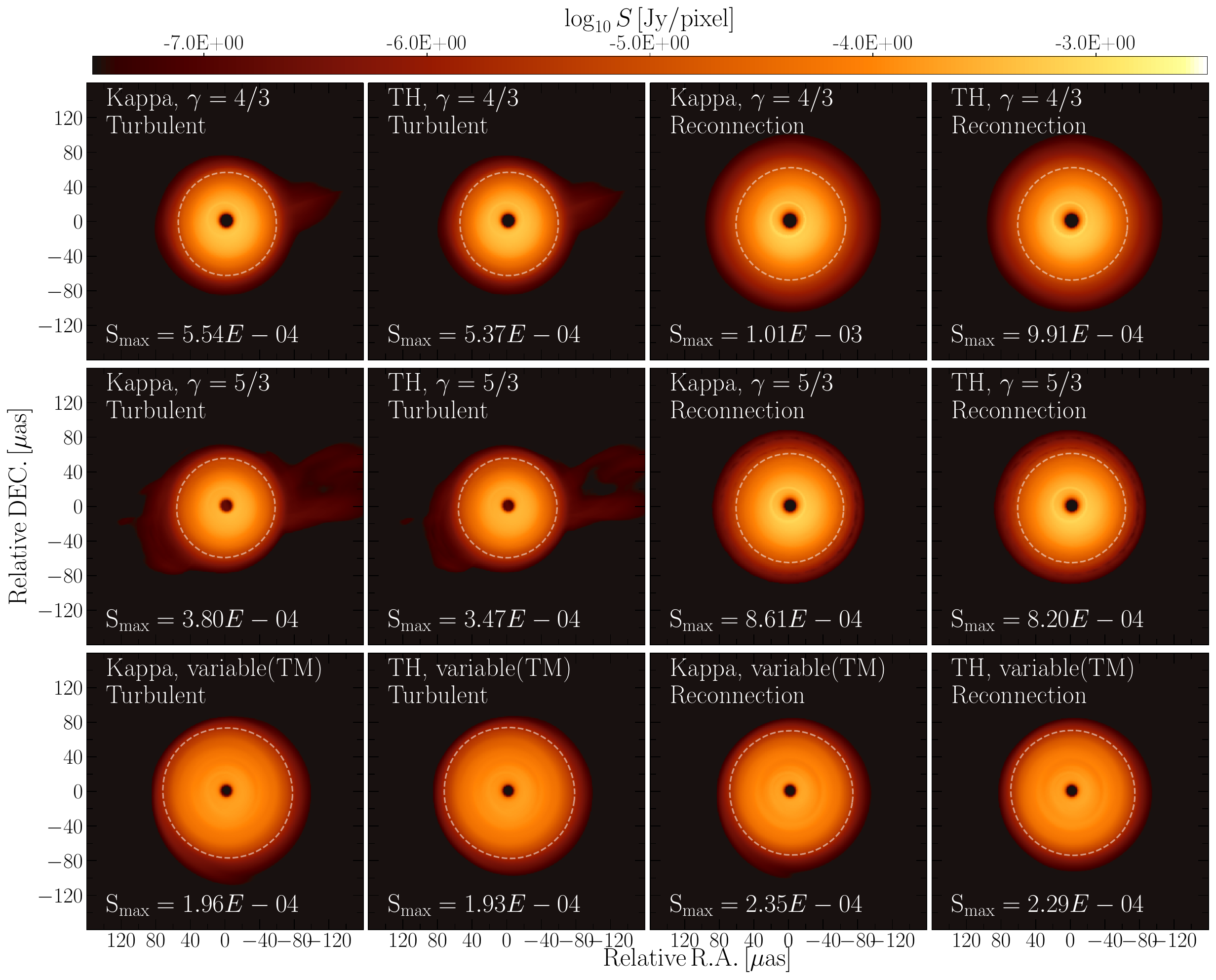}
  \caption{Same as Fig.~\ref{mad-86} but shown for SANE models.}
  \label{sane-86}
\end{figure*}

Here we show the same for the SANE models in Fig.~\ref{sane-230} ($230$GHz) and Fig.~\ref{sane-86} ($86$GHz).
We note that the SANE models use a relatively compact initial torus size. As discussed by \citet{Fromm-etal2022}, such compact SANE torus setups can affect the extended 86 GHz morphology, since lower-frequency emission probes the diffuse extended emission from larger radii and may remain sensitive to the amount of material supplied by the initial torus. This may influence the apparent radial extent, smoothness, and low-surface-brightness structure of the 86 GHz images. In the present work, we nevertheless keep this compact torus setup in order to use a similar initial model for the EoS comparison. Therefore, the SANE 86 GHz morphology should be interpreted with this limitation in mind.

We quantify the influence of the compact initial torus directly. By $t = 13000-15000\,t_{\rm g}$, about 30\% of the initial torus mass has been accreted in the TM SANE model, so the accretion flow is continuously fed rather than exhausted during the analysis window. Moreover, the 1\%-of-maximum intensity contours of the time-averaged images (Figs.~\ref{sane-230} and \ref{sane-86}) have radii of $\simeq 40$--$48\,\mu$as ($\simeq 10.6$--$12.6\,r_g$) at 230~GHz and $\simeq 57$--$76\,\mu$as ($\simeq 15$--$20\,r_g$) at 86~GHz, showing that the reported morphologies are dominated by the turbulent inner inflow; emission from larger radii, which may remain sensitive to the initial torus, contributes only below the 1\% surface-brightness level. We therefore interpret the extended SANE 86~GHz morphology with this caveat in mind.

In general, the SANE models show trends that are qualitatively similar to those of the MAD cases, but there are some quantitative differences in the structure and extent of the emissions. The emission is generally less compact and more disk-dominated, which is a sign of the weaker magnetization and less jet activity in SANE flows. 

In the present simulations, changing the EoS produces larger differences in the brightness distribution and emission extent than changing the electron-heating prescription. These differences should be interpreted as systematic consequences of full EoS-dependent evolution including thermodynamic closure and torus structure.

At 86 GHz, the differences between thermal and $\kappa$ electron distributions are more obvious. This is because non-thermal effects make emission stronger at larger radii, which makes intensity profiles smoother and larger. Nonetheless, similar to the MAD models, variations from electron-heating prescriptions are still subordinate to those resulting from the EoSs. These results support the conclusions in the main text, indicating that the effects of plasma microphysics on observable signatures are not limited to strongly magnetized (MAD) systems and hold across different accretion states.

\section{Central Intensity Profiles}
\label{appendixB}

\begin{figure*}[ht]
    \centering
    \includegraphics[width=0.99\linewidth]{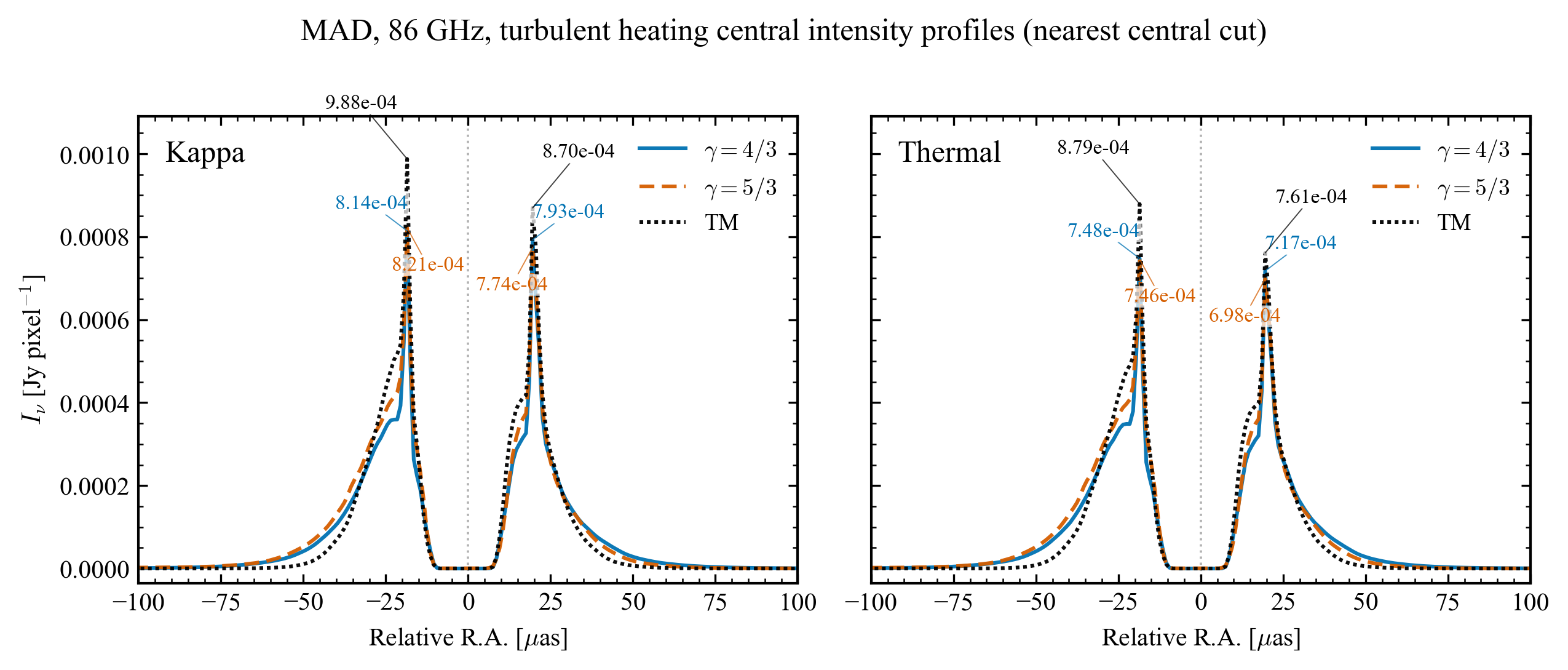}
    \includegraphics[width=0.99\linewidth]{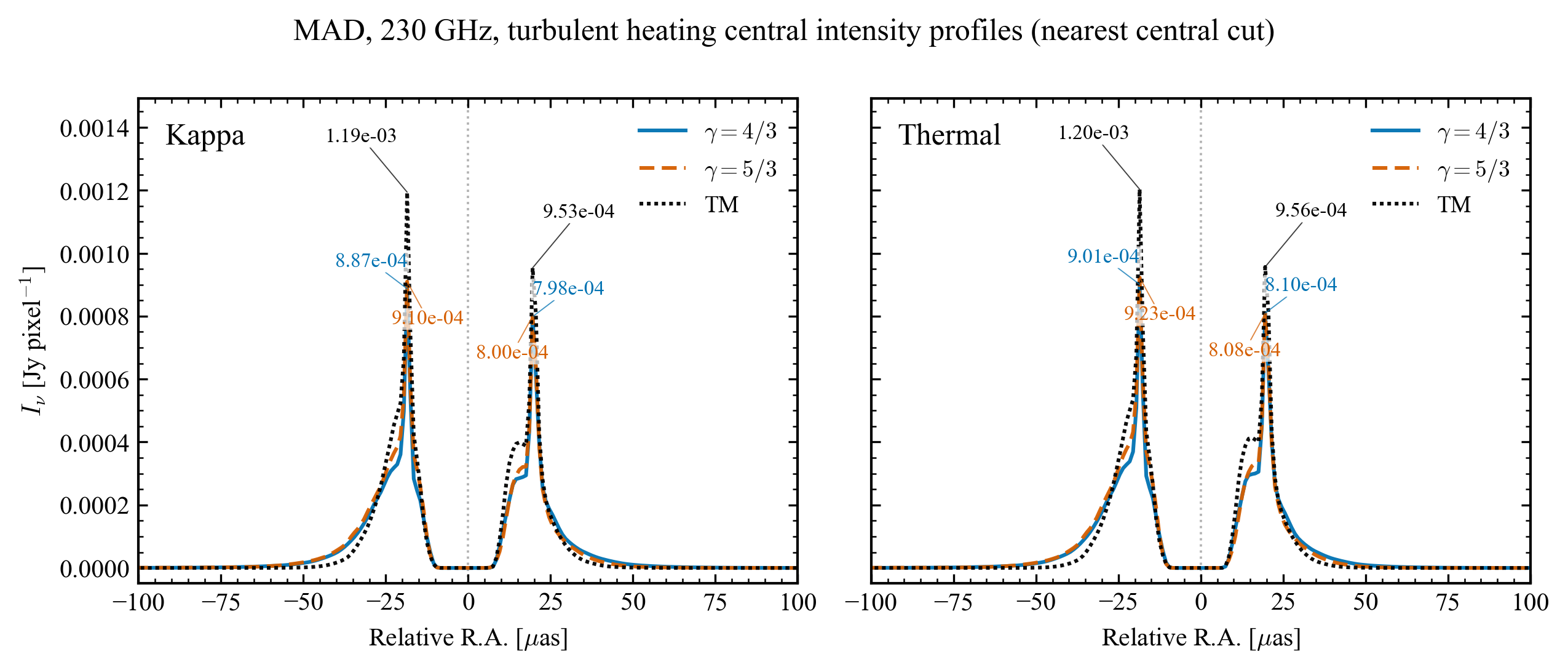}
    \caption{Central horizontal intensity profiles for the turbulent-heating models at 86~GHz (top) and 230~GHz (bottom). The profiles are extracted from the time-averaged images along the cut closest to the image midplane. Each panel compares the three EoS prescriptions for the \(\kappa\) and thermal electron distribution functions, with peak intensity values marked near the corresponding maxima.}
    \label{fig:flux-cut}
\end{figure*}

To further quantify the image morphology discussed in Section~\ref{results03}, we show in Fig.~\ref{fig:flux-cut} the one-dimensional linear intensity profiles at $y\approx0$ extracted from the time-averaged images in Figs.~\ref{mad-230} and~\ref{mad-86}. The profiles are measured along the image center and are shown for both 230 GHz and 86 GHz using the turbulent heating prescription to compare how intensity is distributed along the x-axis under three different EoSs.

These cuts provide a more quantitative comparison of the compact emission region. Since the magnetic reconnection heating prescription shows the same qualitative trends as the turbulent heating case, we show only the turbulent heating results here. 
In both frequencies, the TM models generally show $\approx10\%$ higher peak intensities and $\approx25\%$ narrower concentration of emission around the central bright structure than the constant-\(\gamma\) EoS cases. This supports the qualitative impression from the images that the TM case produces a brighter and more centrally concentrated compact region.

\section{Radial mass-flux structure of the MAD simulations}
\label{appendixC}

To examine whether the inner regions of the MAD simulations have approached a quasi-steady inflow state, we calculate the time-averaged vertically shell-integrated radial mass flux during our simulation time from $t=13000-15000\,M$. Following the diagnostic adopted in Appendix~D of \cite{RN6}, we evaluate
\begin{equation}
\langle  F_{\rm M}(r)\rangle_t 
=
\left \langle -\int_{0}^{2\pi}\int_{0}^{\pi}
\sqrt{-g}\,\rho u^{r}\,d\theta\,d\phi \right \rangle_t ,
\end{equation}
where $g$ is the metric determinant, $\rho$ is the rest-mass density,
and $u^{r}$ is the radial component of the four-velocity.

\begin{figure*}[ht]
    \centering
        \includegraphics[width=0.7\linewidth]{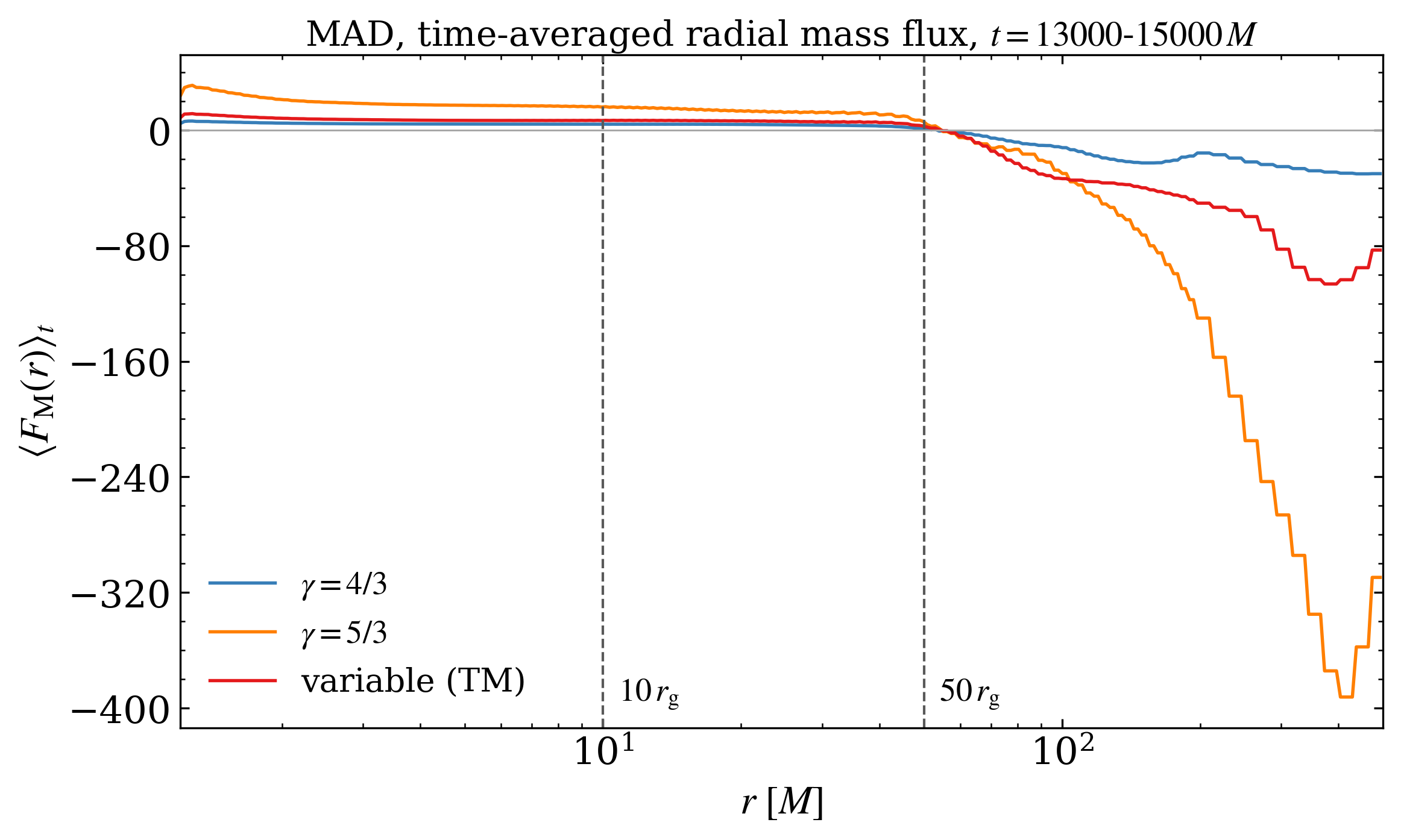}
  \caption{Time-averaged, vertically integrated radial mass accretion flux for the three EoS models over $t=13000$-$15000\,M$ under the MAD state. The vertical dashed lines mark $r=10\,M$ and $50\,M$.}
  \label{f16}
\end{figure*}

Fig.~\ref{f16} presents the profiles for the three EoSs for MAD models. A weak radial dependence of $F_{\rm M}$ is expected in a region approaching inflow equilibrium. Compared with the large variations at larger radii, the profiles are substantially flatter in the inner region ($r\lesssim50\,r_g$) relevant to the horizon-scale emission, supporting an approximately quasi-steady inner flow at the analysis time in addition to Fig.~\ref{fig:MADflow}. We have already noted in the main text that the SANE models reach a quasi-steady state earlier than the MAD models. We therefore analyze them over the same late-time interval but do not show a separate radial mass-flux figure to avoid repetition. For the MAD models, the approximate flatness of $F_{\rm M}$ inside $r\lesssim50\,r_g$ indicates that the mass supply to the inner flow is approximately steady while the images are computed, which implies that the reported EoS-dependent differences are unlikely to be dominated by secular draining of the initial torus. The next section, Appendix~\ref{appendixD}, compares the initial setup for the different models.

\section{Torus Geometry and Vertical Structure}
\label{appendixD}

\begin{figure}
    \centering
    \includegraphics[width=0.45\linewidth]{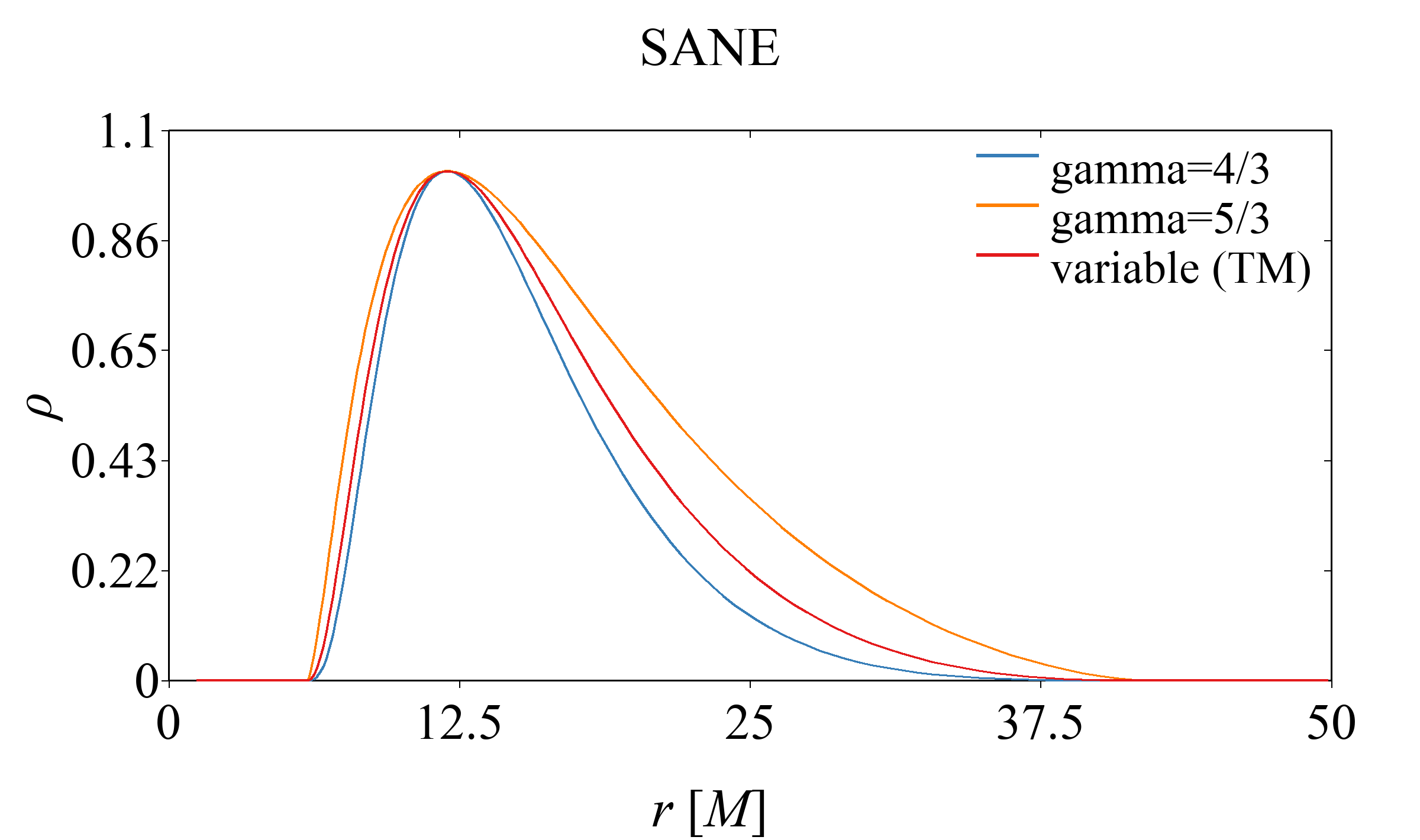}
    \includegraphics[width=0.45\linewidth]{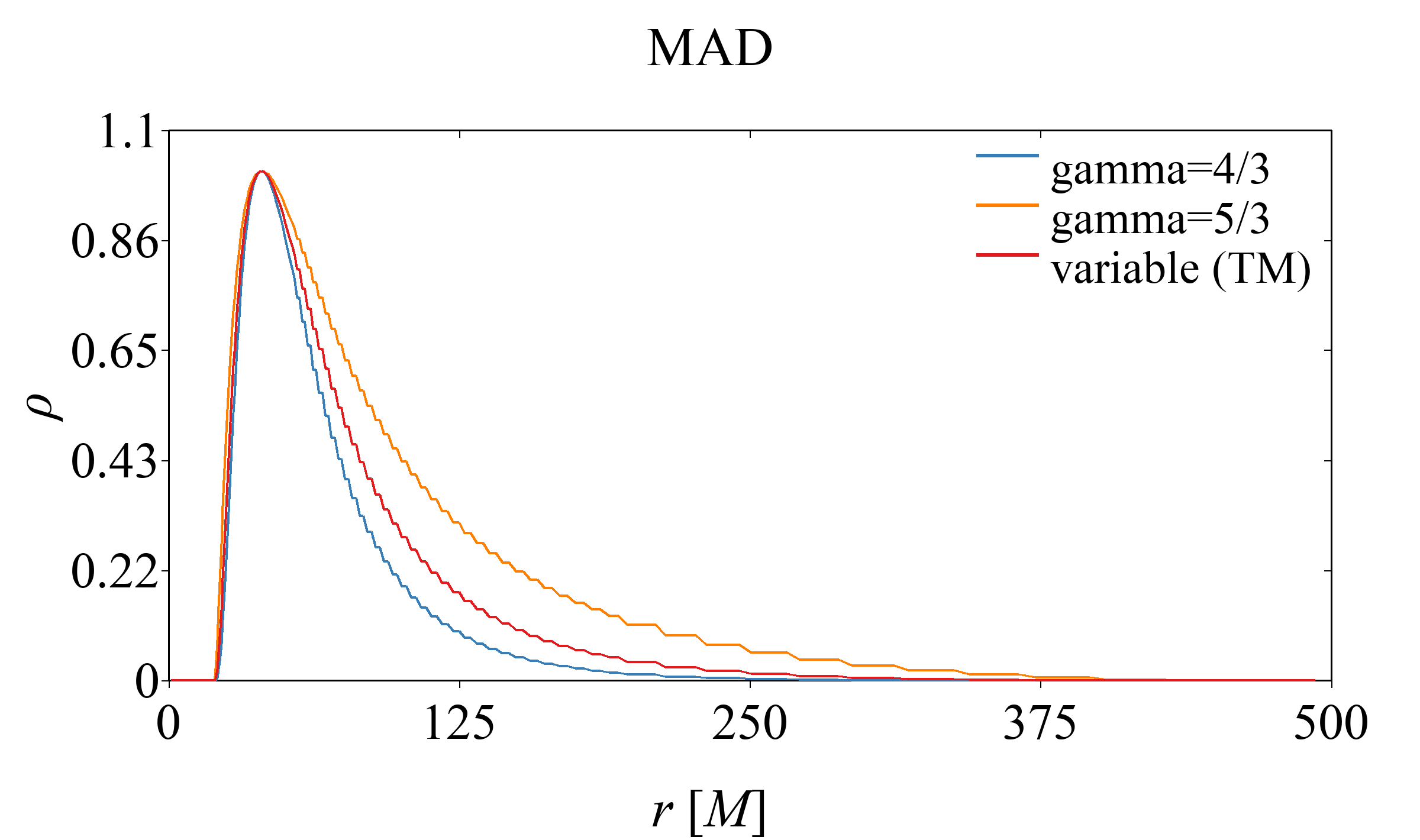}
    \caption{Equatorial rest-mass density profiles of the initial SANE (left) and MAD (right) tori for the different EoS cases. The blue, orange, and red curves correspond to $\gamma=4/3$, $\gamma=5/3$, and the variable TM EoS, respectively. }
    \label{fig:initial-torus}
\end{figure}

Because the adiabatic index is involved in the calculations of Fishbone-Moncrief hydrostatic equilibrium, the initial tori are not necessarily identical even when their inner edge, density maximum, and peak normalization are fixed. We therefore quantify the initial and evolved torus geometry before interpreting the differences found later in the simulations. Fig.~\ref{fig:initial-torus} shows the equatorial rest-mass density profiles of the initial tori for SANE (left) and MAD (right) models. Although all models have the same prescribed inner edge and density maximum radius within each accretion state, their radial density profiles differ slightly away from the maximum because of the different adiabatic indices of the hydrostatic equilibrium solution of the FM torus.

\begin{figure}
    \centering
    \includegraphics[width=0.8\linewidth]{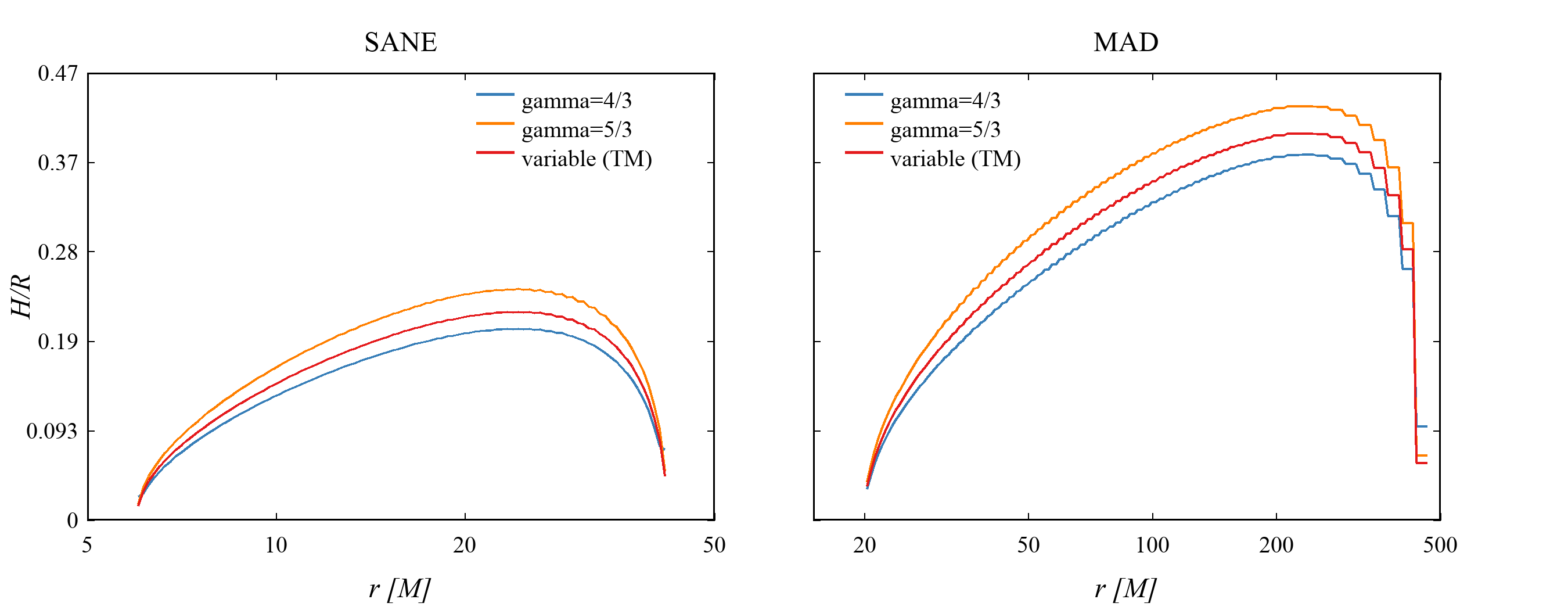}
    \caption{Radial profiles of the initial disk aspect ratio, $H/R$, for the SANE (left) and MAD (right) tori constructed with the three EoS prescriptions. The blue, orange, and red curves correspond to $\gamma=4/3$, $\gamma=5/3$, and the variable TM EoS, respectively.}
    \label{initial-HR}
\end{figure}

The corresponding initial spherical scale heights are shown in Fig.~\ref{initial-HR}, where
\begin{equation}
\left \langle \frac{H}{R} \right\rangle_t
=
\tan\left(
\frac{
\displaystyle \left \langle \int
\left|\theta-\pi/2\right|
\rho\sqrt{-g}\,{\rm d}\theta\,{\rm d}\phi \right\rangle_t
}{
\displaystyle \left \langle \int
\rho\sqrt{-g}\,{\rm d}\theta\,{\rm d}\phi \right\rangle_t
}
\right).
\label{eq:disk-height}
\end{equation}
This diagnostic provides a measure of the density-weighted vertical thickness of the torus. The TM case generally lies between the two constant-$\gamma$ cases, while the constant-$\gamma=5/3$ and $\gamma=4/3$ models bracket the thicker and thinner configurations, respectively. Because the SANE torus is more compact, these initial geometric differences are potentially more relevant for its subsequent evolution than for the larger MAD torus.


\begin{figure}
    \centering
    \includegraphics[width=0.7\linewidth]{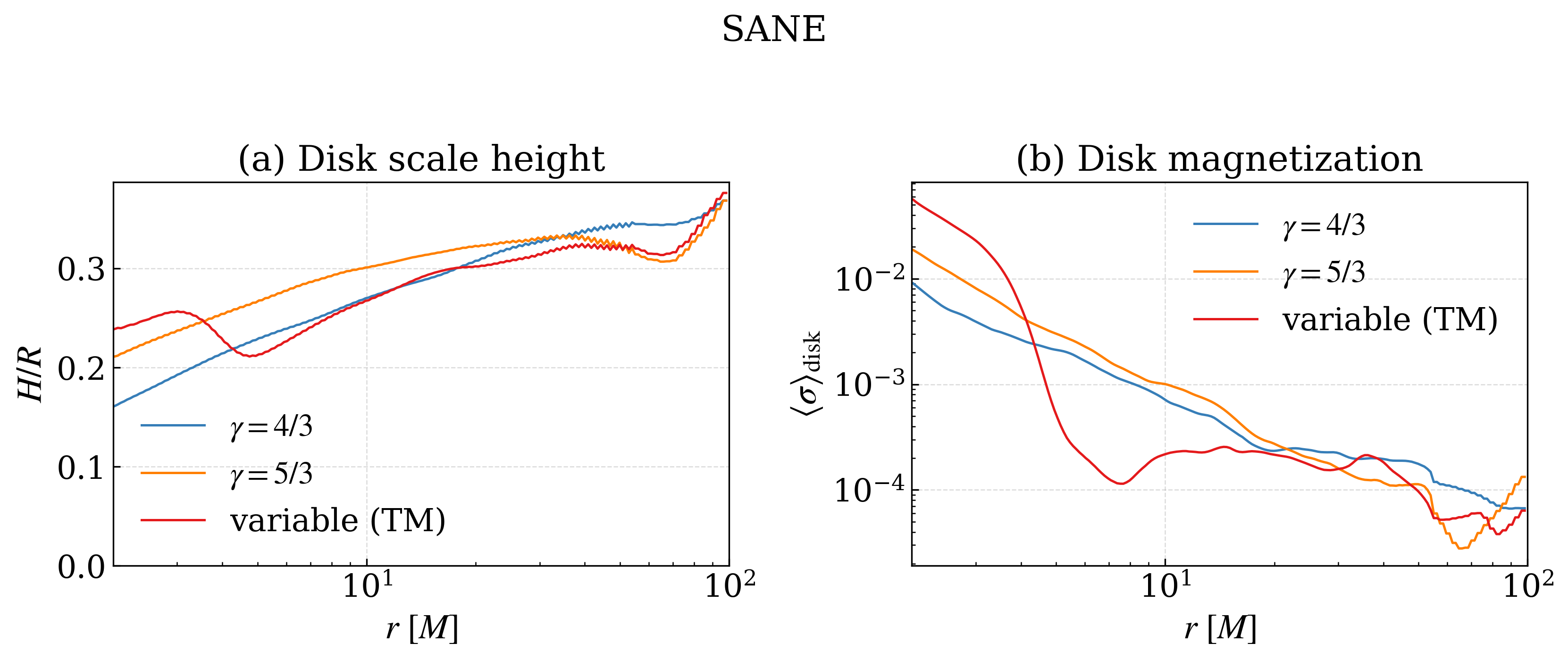}
    \includegraphics[width=0.7\linewidth]{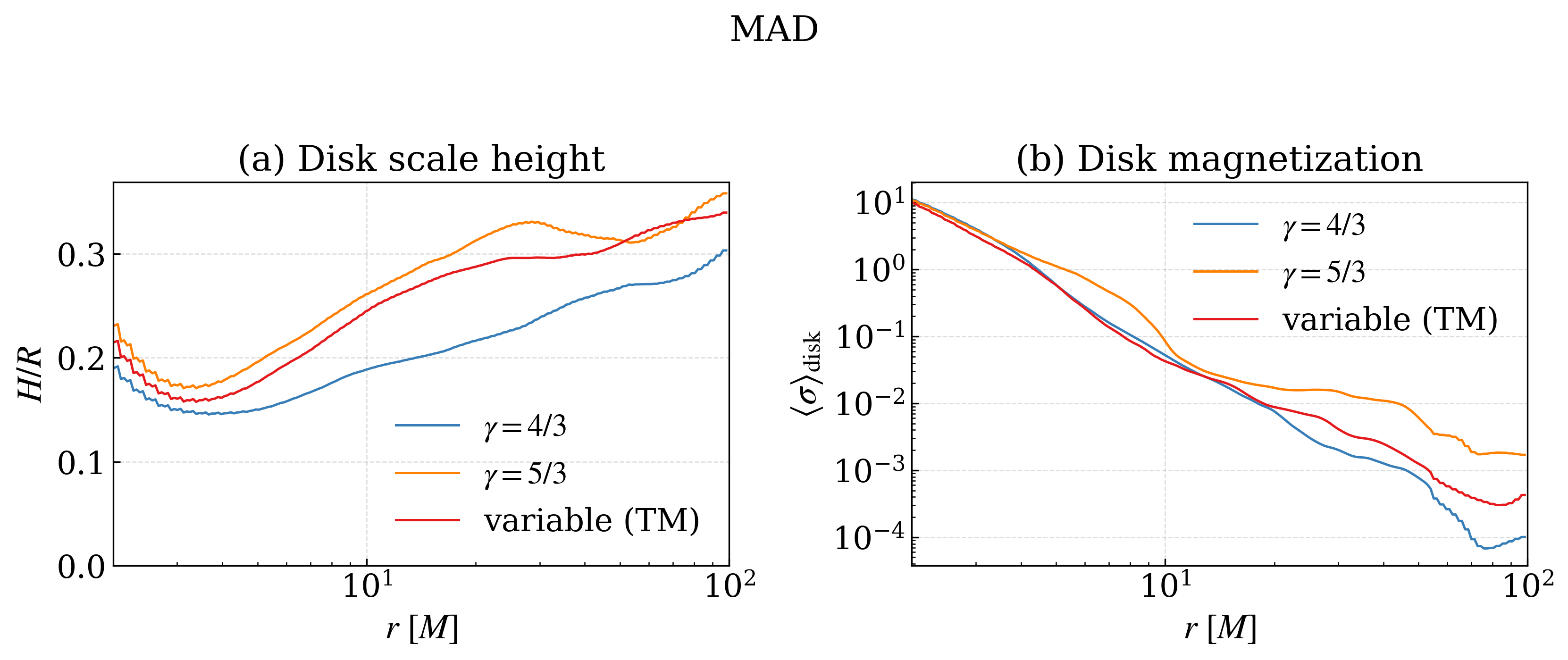}
    \caption{Time-averaged radial profiles of the disk scale height and magnetization for the SANE (top) and MAD (bottom) cases over $t=13000$--$15000\,t_{\rm g}$. The left panels show $H/R$, while the right panels show the disk-averaged magnetization, $\langle\sigma\rangle_{\rm disk}$, on a logarithmic scale.}
    \label{HR-sigma}
\end{figure}

Fig.~\ref{HR-sigma} presents the spherical scale height and disk-averaged magnetization, \(\sigma=b^2/\rho\), averaged over \(t=13000\)-\(15000\,M\). The disk properties after time-evolution exhibit EoS-dependent differences in both quantities, indicating that changing the thermodynamic closure affects not only the local temperature but also the global disk geometry and magnetic structure. The SANE models show particularly noticeable differences in the inner radial profiles, whereas the MAD models preserve broadly similar radial trends but differ quantitatively in disk thickness and magnetization.

These results are qualitatively consistent with \cite{White_2020}, who found that radiatively inefficient accretion flows in inflow equilibrium do not necessarily converge to a universal radial structure and may retain sensitivity to the EoS and the initial magnetic-field configuration.

\begin{figure}
    \centering
    \includegraphics[width=0.85\linewidth]{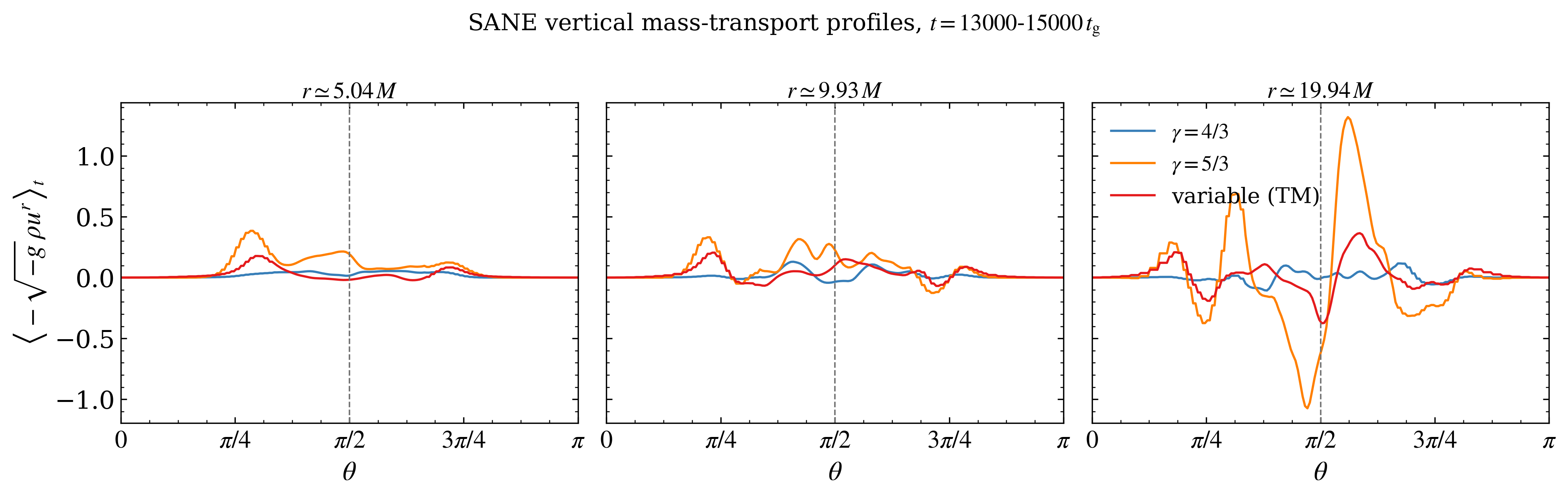}
    \caption{Time-averaged vertical profiles of the radial mass flux,$\langle-\sqrt{-g}\rho u^r\rangle_t$, at $r\simeq5$, $10$, and $20\,r_g$ for the three SANE EoSs over $t=13000$-$15000\,t_{\rm g}$. The dashed lines mark the disk midplane, $\theta=\pi/2$.}
    \label{sane-vertical}
\end{figure}

We further examine the vertically resolved radial mass flux at
\(r=5\), \(10\), and \(20\,r_g\) in
Fig.~\ref{sane-vertical}.
The mass transport is vertically structured and varies among the EoS models, showing that EoS-dependent disk geometry can influence accretion and magnetic-flux transport, which should especially be considered in the SANE state. Consequently, the differences reported in this work should be interpreted as systematic consequences of changing the thermodynamic closure, including both direct thermodynamic effects and geometry-mediated dynamical effects. The present simulations do not allow these two contributions to be separated uniquely.

\bibliography{ref2}{}
\bibliographystyle{aasjournal}

\end{document}